\documentclass[AMA,STIX1COL]{WileyNJD-v2}

\usepackage{verbatim}
\usepackage{cancel}
\usepackage{url}

\usepackage{dblfloatfix} 
\usepackage{natbib}

\usepackage{enumitem}
\usepackage{adjustbox}
\usepackage{xcolor}
\usepackage{amssymb}
\usepackage{amsbsy}
\usepackage{algorithm,algorithmicx}
\usepackage{algpseudocode}
\setcitestyle{super} % Leandro: this seems to be format in published papers
\algnewcommand{\LeftComment}[1]{\Statex \(\triangleright\) #1}

\makeatletter
\renewcommand{\ALG@name}{Transaction code}
\makeatother

\usepackage{caption}
\usepackage{subcaption}
\usepackage{amsmath}

\newlist{myenumi}{description}{10}
\setlist[myenumi]{labelindent=\parindent, leftmargin=*, label=(\roman*), align=left}
\setlist[myenumi]{leftmargin=0pt}

\graphicspath{{./images/}}
\articletype{Paper Submission}%

\everypar{\looseness=-1}

\begin{document}

\title{Blockchain for Economically Sustainable Wireless Mesh Networks}

\author[1]{Aniruddh Rao Kabbinale}

\author[2,3]{Emmanouil Dimogerontakis}

\author[2,3]{Mennan Selimi}

\author[1]{Anwaar Ali}

\author[2,3]{Leandro Navarro}

\author[3]{Arjuna Sathiaseelan}

\author[1]{Jon Crowcroft}
\authormark{A R Kabbinale \textsc{et al}}

\address[1]{University of Cambridge, Cambridge, UK}

\address[2]{Universitat Polit\`{e}cnica de Catalunya. Barcelona, Spain}

\address[3]{Ammbr Research Labs, Cambridge, UK}

\corres{Aniruddh Rao Kabbinale \\\email{aniruddh@maargin.in}}

\presentaddress{198, MS Ramaiah City Layout, JP Nagar 8th Phase, Bengaluru, India - 560078}
%\address[3]{[a1,a2,a3]@ac.upc.edu}
%\address[4]{a4@cl.cam.uk}

%\address[3]{\orgdiv{Org Division}, \orgname{Org Name}, \orgaddress{\state{State name}, \country{Country name}}}

%\corres{Mennan Selimi, Edifici C6-E208 C.Jordi Girona, 1-3 08034 Barcelona, Spain. \email{mselimi@ac.upc.edu}}

%\presentaddress{This is sample for present address text this is sample for present address text}

\abstract[Summary]{Decentralization, in the form of mesh networking and blockchain, two promising technologies, is coming to the telecommunications industry. Mesh networking allows wider low cost Internet access with infrastructures built from routers contributed by diverse owners, while blockchain enables transparency and accountability for investments, revenue or other forms of economic compensations from sharing of network traffic, content and services. Crowdsourcing network coverage, combined with crowdfunding costs, can create economically sustainable yet decentralized Internet access. This means every participant can invest in resources, and pay or be paid for usage to recover the costs of network devices and maintenance. While mesh networks and mesh routing protocols enable self-organized networks that expand organically, cryptocurrencies and smart contracts enable the economic coordination among network providers and consumers. We explore and evaluate two existing blockchain software stacks, Hyperledger Fabric (HLF) and Ethereum geth with Proof of Authority (PoA) intended as a local lightweight distributed ledger, deployed in a real city-wide production mesh network and also in laboratory network. We quantify the performance, bottlenecks and identify the current limitations and opportunities for improvement to serve locally the needs of wireless mesh networks, without the privacy and economic cost of relying on public blockchains. 
%The results show that both Hyperledger Fabric and geth Ethereum network can be deployed on even resource constrained devices like RPI3 boards or router boards with limited computational capability. Both the blockchain software stacks perform well without saturation and much delays for a moderate load of up to 100 transactions fired in the network at a time. In Hyperledger Fabric, our measurements reveal that endorsers are the bottleneck and care has to be taken in designing endorsement policy for scaling the network. In case of Ethereum, our results show that a there is a limit on the number of requests a node can support and can only be scaled vertically i.e. by increasing computational capability of serving node. %\lnm{Say something about the main results ...}
% To the best of our knowledge, this is the first HLF deployment made in a production wireless mesh network. The goal of this paper is to conduct a performance analysis of two popular blockchain platforms  in private setting, Hyperledger Fabric and Ethereum, assess the performance and limitations of these two state-af-the-art platforms in a production wireless mesh network. \mennan{Inform practitioners in making decisions regarding adoption of blockchain technology in wireless networks ?}
}
 

%\jnlcitation{\cname{%
%\author{Williams K.}, 
%\author{B. Hoskins}, 
%\author{R. Lee}, 
%\author{G. Masato}, and 
%\author{T. Woollings}} (\cyear{2016}), 
%\ctitle{A regime analysis of Atlantic winter jet variability applied to evaluate HadGEM3-GC2}, %\cjournal{Q.J.R. Meteorol. Soc.}, \cvol{2017;00:1--6}.}

\maketitle

\section{Introduction}

%\todo[inline]{Leandro: clarify terminology: either Wireless Mesh Networks or Community Mesh Networks. I think no need to overcomplicate the acronmy with WCMN, CMN should be enough. We can just talk about WMN (as in the title) and use the case of guifi.net and CMN as the scenario we deploy/test it, and therefore make the paper more general. What do you think? You can remove when Ack'd. \todo[inline, color=green!40]{AA: I have tried to address this comment. I will remove this todo block if the updated text seems fine to everyone.}}

Network infrastructures are critical to provide local and global connectivity that enables access to information, social inclusion and participation for everyone. Local connectivity largely relies on access networks. 
Wireless mesh networks (WMNs) are a kind of access networks comprising of wireless nodes namely wireless mesh routers, wireless mesh clients, and network gateways. A client (connected through WiFi or wired to a mesh router) can access the Internet across a WMN \citep{akyildiz2005wireless}. These are self-organized networks that can grow organically: new network links can expand the coverage of the network or increase the capacity when links get overused. The routing protocol runs in every router by measuring the performance and quality of links and coordinates distributed decisions about the best network paths periodically. As a result, once a routing protocol is adopted, the development and operation of the network only depends on pooling routers and links with local decisions, without any central planning or management.

These decentralized networks are essential to develop community access networks, network infrastructure commons, built by citizens and organizations which pool their resources and  coordinate their efforts, characterized by being open, free and neutral~\citep{baig2016making}.
%
% \lnm{redundant} Community Mesh Networks (CMNs) are a special case of WMNs which are usually setup as a community network.
%where the access network is built, owned, and operated in a peer-to-peer and collaborative manner by various stakeholders in a community (such as citizens of a geographic region with poor Internet access) pooling their resources (routers and links).
These decentralized access networks have been identified as one way to connect the next billion people that are still without the Internet access \citep{ITU}. Guifi.net\footnote{\url{https://guifi.net/}} is an example of such a community effort which is one of the biggest community networks in the world, with more than $34,000$ participating routers, combining technologies including wireless mesh and fibre. However the main challenge of these  peer-to-peer socio-technical structures are around trust among agreements between peers and how to ensure the economic sustainability of this collective effort and the balance between contribution and consumption~\citep{baig2016making}.

%\textbf{Example: Compensation System in Guifi.net. How it is envisioned for the moment ? How it is implemented ? Is it automated ? If yes, is it trusted automation ? }

An example scenario and mechanism for economic sustainability is the economic \emph{compensation system} used in Guifi.net \citep{baig2016making}. An answer to the lack of incentives to invest in network infrastructure, it was introduced in 2011 as a cost sharing mechanism. The idea of the compensation system is to balance between total resource contribution and its consumption. The economic cost of any contribution and consumption of network resources by each participant in a given locality are recorded. The overall result is a zero-sum computed periodically, from monthly to quarterly, where the participants with over-consumption or negative balances have to compensate those with over-contribution or positive balances. 
%If, for instance, a participant consumes more than it contributes then its balance goes into negative and this participant is, in turn, charged an amount based on this difference. If, however, a participant's consumption is relatively less as compared to how much it contributes to the network then the WCMN itself owes some monetary amount to such a participant.

Currently the above described economic compensation system is done manually: each participant declares its costs and consumption and then the Guifi.net foundation\footnote{\url{https://fundacio.guifi.net/Foundation}} validates this claim by cross checking it with their own network traffic measurement data and network inventory, according to the agreed list of standard costs. Any disparities between these two records are flagged, clarified or raised to a conflict resolution mechanism. There is, however, room for error or intentional false or exaggerated claims put forward by a participant, the recorded data being tampered with, or simply mistrust among the parties. The correct application of the compensation system is critical for the economic sustainability of the network, ensuring its proper operation, as well as future investments. Therefore, we argue that there is a need for an automated system where diverse participants, resource providers and consumers, can trust that the consumption of resources is being accounted in a fair manner, and that these calculations and money transfers are automated, irreversible and shared across different participants, to avoid the cost, delays, errors and potential mistrust from manual accounting and external payments. %this leads to need for a fair and transparent compensation system.

%\todo[inline, color=green!40]{Leandro: I can add more details about how the compensation system can help the econ sustainability of the mesh: payment proportional to usage, availability, etc. Probably enough as is. You can remove when clear.}
%The participants who use a significant amount of resources from the CPR are obliged to sign an agreement for economic activities and for the participation in the compensation system

%\textbf{After talking to Leandro:} At the moment the Guifi Foundation (GF) is doing a network monitoring (traffic measurement) for the Compensation System (CS). This process is totally manual (done in a website): Each user (or ISP) uploads the proofs for compensation, then the GF downloads these form and computes the compensation. Both parties need to agree on the values (both parties measure the traffic). However, this process is a very slow process and not accurate (done every month) which will not be the case in Ammbr. 

%\manos{The message we want to transmit is that trusted automation of the compensation system + accountability can lead to economical sustainability}
Blockchain technology offers solutions that seems apt to make the peer-to-peer nature of access networks trusted and economically sustainable. Blockchain (more details in Section \ref{sec:bc}) is an immutable and distributed data storage without the provision of retrospective mutation in data records. However, most blockchain networks are open and public (permissionless) that encourage the users to protect anonymity \citep{nakamoto2008bitcoin}. This implies that anyone, without revealing their true identity, can be part of such a network and make transactions with another similarly pseudonym peer of the network.

In the perspective of community networks such as Guifi.net, however, every participant who joins the network to contribute and benefit from the infrastructure must first register its identity and the identity of the resources that it contributes to the wider pool. This is particularly needed so that any malicious entity, such as hidden nodes in Guifi.net used by other ISPs, can be filtered out \citep{Neumann:2016}. Because of such registration process one also needs an efficient identity mechanism on top of blockchain's immutable record keeping. \emph{Permissioned blockhains} are part of such solutions, mostly envisioned for business networks where there is often a stringent requirement of \emph{know your customer} in addition to keeping the intra- and inter-business transactions confidential. 

In this study, we extend our previous work \cite{Selimi2018CryBlock} by exploring the plausibility of combining decentralized access networks with a permissioned blockchain running on servers inside the access network, that would result in a model for economically self-sustainable decentralized mesh access networks, guaranteeing trust among participants, allowing economic profitability, and enabling at the same time easier Internet connectivity. We study the viability of such an approach, by evaluating two of the most prominent platforms for building local blockchain applications. %\manos{we can do maybe beforehand a separation between blockchains that are tied to a concrete product like Filecoin, Bitcoin etc and blockchains that offer an environment for creating applications}. 
These platforms are Hyperledger Fabric (HLF)\footnote{\url{https://www.hyperledger.org/projects/fabric}}, an industry-oriented modular, and permissioned distributed ledger and the Ethereum\footnote{\url{https://www.ethereum.org/}}, a general-purpose platform acting as a permissioned blockchain through the lighter PoA validators among a small set of replicas inside the network. 
%is one such solution that realizes the concept of permissioned blockchains and one which we also use in our current study.

We deploy the Hyperledger Fabric and Ethereum PoA platform in a local network in our laboratory, and well as in a decentralized production wireless mesh network that is part of Guifi.net. 
Our key contributions are summarized as follows:
\begin{itemize}
    \item First, we analyze the performance of both platforms in terms of metrics such as transaction latency, CPU and memory utilization of Hyperledger Fabric and Ethereum PoA components. To the best of our knowledge, this is the first Hyperledger Fabric and Ethereum PoA deployment
    made in a production wireless mesh network. %\mennan{Quantification of throughput and latency in HLF and Ethereum would/can enable practitioners to have have insight understandings of performance and limitations of existing blockchain technologies in wireless mesh networks.}
   Our results show that both Hyperledger Fabric and geth Ethereum PoA network can be deployed on even resource constrained devices like RPI3 boards or router boards with limited computational capability. Both the blockchain software stacks perform well without saturation and much delays for a moderate load of up to 100 transactions fired in the network at a time. In Hyperledger Fabric, our measurements reveal that endorsers are the bottleneck and care has to be taken in designing endorsement policy for scaling the network. In case of Ethereum PoA, our results show that a there is a limit on the number of requests a node can support and can only be scaled vertically i.e. by increasing computational capability of serving node.

    \item Second, driven by the findings in a mesh network, we propose a placement scheme for Hyperledger Fabric and Ethereum PoA components that optimizes the performance of the blockchain components in mesh networks.
\end{itemize}

The rest of the paper is organized as follows. In Section~\ref{sec:bc}
 we briefly discuss the target blockchain platforms, Hyperledger Fabrics and Ethereum, and the way how their protocol works. In Section~\ref{sec:qmp} we describe and characterize the performance of the QMPSU mesh network and testbed where our experiments are performed. In Section~\ref{sec:eval} the performance of Hyperledger Fabrics and Ethereum platform is presented and our main findings are discussed. Section \ref{sec:relwork} describes related work and section \ref{sec:conclusion} concludes and discusses future research directions.

 % Introduction

\section{Blockchain: The underpinning Technology}
\label{sec:bc}

%Blockchain is a distributed ledger that consists of blocks of transactions and their corresponding hashes chained together providing immutability. By storing blocks of information that are identical across its network, the Blockchain cannot: i) Be controlled by any single entity ii) Has no single point of failure. Blockchain technology has found significant attention in recent times due to the widening popularity of BitCoin [ref] – the first Blockchain based cryptocurrency. BitCoin using Blockchain technology to provide a peer-to-peer decentralised crypto currency based transactions. The underlying foundations of Blockchain are based on cryptography and consensus mechanisms to establish security and ensure both validity and consistency of the transactions.

Blockchain is an \emph{append-only} immutable data structure. Its first incarnation was in the Bitcoin cryptocurrency network \citep{nakamoto2008bitcoin}. Blockchain was used to enable trust in financial transactions among different non-trusting parties in a pure peer-to-peer fashion without the need for going through a third financial party like e.g., a bank. Such trust is provided in terms of immutability of blockchain's data structure. Each \emph{block} in blockchain contains information that is immutable. The immutability aspect is rendered true by including the hash of all the contents of a block into the next block which also chains the blocks together. Tampering with one block disturbs the contents of all the following blocks in the chain. Each block in the chain is appended after a \emph{consensus} is reached among all the peers of the network. The same version of a blockchain is stored in a distributed manner at all the peers of the network. That is why it is sometimes referred to as \emph{distributed ledger} as well.

In this section, we briefly discuss two blockchain platforms chosen for evaluation of economic compensations among consumers and providers of connectivity in wireless mesh networks, as automated calculations with irreversible transactions and money transfers, shared across several participants. These are Hyperledger Fabric and Ethereum, due to their popularity, maturity and potential to be used in different applications.

\subsection{Permissionless vs Permissioned, Public vs Private}

Bitcoin \citep{nakamoto2008bitcoin} and Ethereum \citep{wood2014ethereum}, as various other blockchains, are considered as \emph{permissionless}, meaning that anyone  has "write" access to the blockchain. As a result anyone can be a part of the network, mining and performing transactions with other parties. The consensus in such an open environment is tackled with algorithms like the Proof-of-Work(PoW) protocol. A potential for anonymity and privacy is also at the heart of such platforms. A user (or in general an entity) usually uses the hash of its public key as pseudonym or a zero-knowledge protocol as opposed to using its real-world name or details. 

In the aspect of "write" openness, \emph{permissioned blockchains} are in sharp contrast with public blockchains which we discuss next.
Permissioned blockchains, a concept particularly popularized by the Linux Foundation's Hyperledger, are usually considered for business applications. In such applications the identity of users, in addition to trusted and immutable data storage, is also important such as the stringent requirement of \emph{know your customers} for many businesses. Hyperledger tries to leverage the best of both worlds by implementing a cryptographic \emph{membership service} on top of blockchain's trusted, immutable, and distributed record keeping. 

Another categorization can be done based on the openness of reading from the blockchain. In the case where a blockchain exposes its data publicly it is characterized as \emph{public}. On the other hand, blockchains that prohibit access to its data are called as \emph{private}.

In our study, the requirement of both users' identity and trusted record keeping is of paramount importance and that is why we decided to conduct our study using private permissioned blockchains. Hyperledger Fabric fulfills by default these properties. On the other hand, while Ethereum is not primarily destined to serve these purposes, it  can also be used as private permissioned blockchain. Nevertheless, executing resource-full consensus algorithms in a permissioned environment where the participants are known has no application except experimentation with the protocols themselves. On the other hand, some protocols, like Ethereum, offer inexpensive consensus algorithms, like the Proof-of-Authority (POA) protocol, that are ideal for a private permissioned instances, as envisaged in our scenario.

\subsection{Hyperledger Fabric (HLF)}
\label{subsec: hlf}

Hyperledger Fabric (HLF) \citep{Hyper2018} is an open source implementation of a permissioned blockchain network that executes distributed applications written in general-purpose programming languages (e.g., Go, Java etc). HLF's approach is modular, which implies that the platform is capable of supporting different implementations of its different components (such as different consensus protocols) in a \emph{plug-and-play} fashion. 

%A permissioned blockchain means that any node is required to maintain a member identity on the network. Even end users must be authorized and authenticated in order to use the network.

The HLF architecture comprises of the following components: 

\begin{description}
    \item[Peers:] Peers can further be of two types namely \emph{endorsers} and \emph{committers}. A peer is called a \emph{committer} when it maintains a local copy of the ledger by committing transactions into its blocks. A peer assumes the role of an \emph{endorser} when it is also responsible for simulating the transactions by executing specific chaincodes and endorsing the result (see the next subsection \ref{subsec: hlfProtocol}). A peer can be an endorser for certain types of transactions and just a committer for others.
    
    \item[Ordering service:] The role of this component is to order the transactions chronologically by time stamping them to avoid the \emph{double spend problem} \citep{nakamoto2008bitcoin}. The ordering service creates new blocks of transactions and broadcast them to the peers which then append these blocks to their local copy of the blockchain (or ledger). The ordering service can be implemented as a centralized or decentralized service \citep{Bessani2017}. It is at the ordering service level where the consensus (like proof-of-work in Bitcoin \citep{nakamoto2008bitcoin}) related to the state of a blockchain takes place.
    
    \item[Chaincode:] A chaincode or a \emph{smart contract} is a program code that implements the application logic. It is run in a distributed manner by the peers. It is installed and instantiated on the network of HLF peer nodes, enabling interaction with the network's shared ledger (i.e., the state of a database modeled as a versioned key/value store). 
    
    \item[Channel:] A channel provides a higher layer of confidentiality abstraction. A channel can be considered as a subnet on top of a larger blockchain network. Each channel has its own set of chaincodes, member entities (peers and orderers), and a distinct version of a distributed ledger. This should not be confused with a similar term, payment channels, used to make multiple off-chain micro-payments, multiple transactions, without committing all to a blockchain. 
    
    \item[Membership service provider (MSP):] HLF makes use of a dedicated and exhaustive Membership Service Provider (MSP)\footnote{\url{https://hyperledger-fabric.readthedocs.io/en/release-1.3/membership/membership.html}}, which is based on public-key infrastructure (PKI) and hierarchical certificate authorities (CAs), to define roles and security clearance (for different channels) of different entities for a particular use case. The goal of such a dedicated MSP is to realize the concept of an organization-like hierarchical security infrastructure in the form of a hierarchical and permissioned version of blockchain.
    
\end{description}

%peer nodes, the Ordering Service and client applications. A peer node can in turn have two roles:  The ordering nodes are responsible for creating blocks for the distributed ledger, as well as the order of transactions in a block to be committed to the ledger. They can  

%The Hyperledger Fabric enables clients to manage transactions by using \emph{chaincodes}. 

\subsubsection{HLF Protocol}
\label{subsec: hlfProtocol}

\begin{figure}[t]
\centering
\includegraphics[width=4.5in,keepaspectratio]{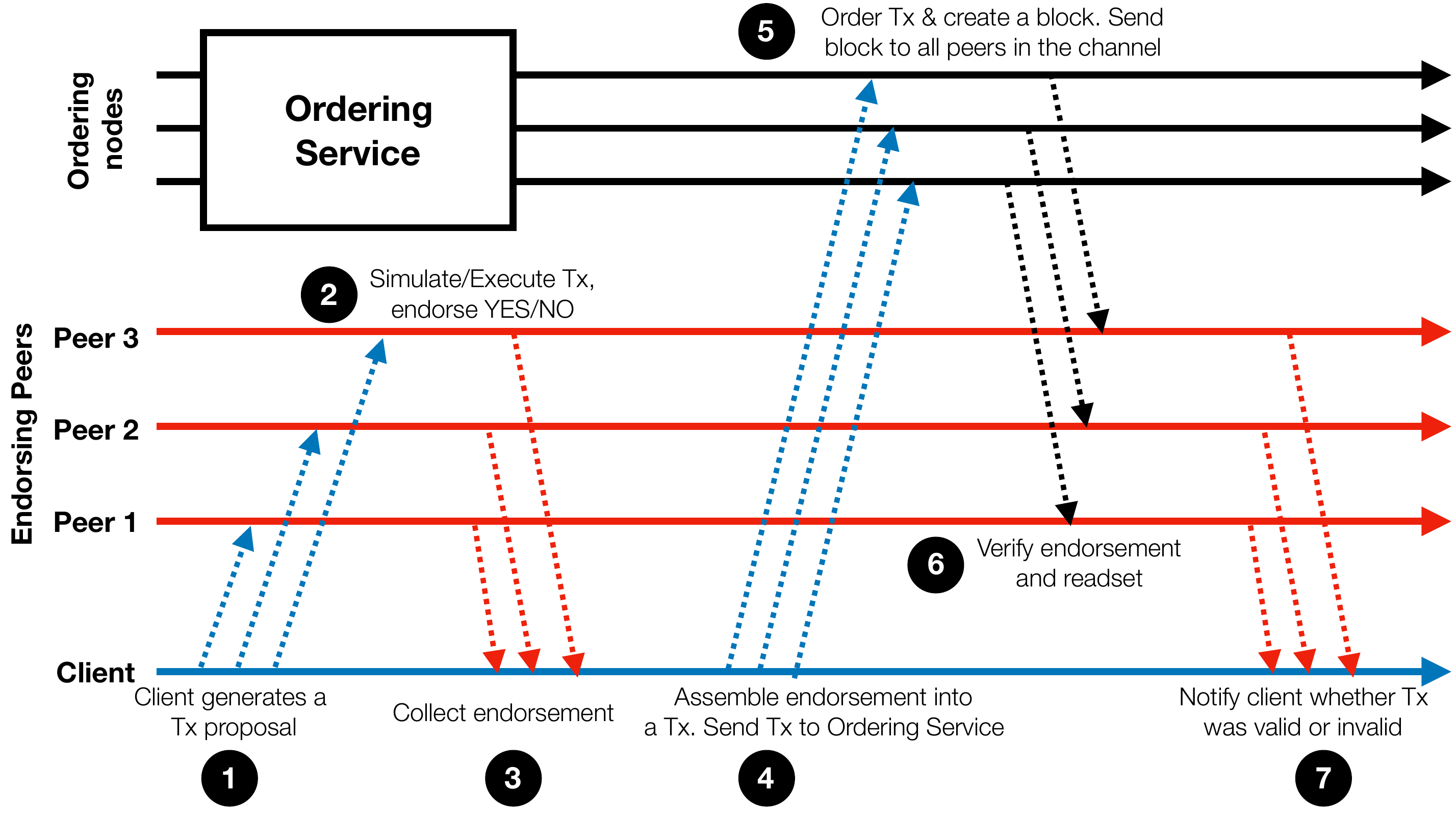}
\captionsetup{justification=centering}
\caption{Hyperledger Fabric Protocol}
\label{fig:hyper}
\end{figure}

Figure \ref{fig:hyper} depicts the sequence of transaction execution steps in HLF's environment. The description of these execution steps are as follows:

\begin{description}
    \item [1. Transaction (Tx) proposal:] In this step clients access the HLF blockchain to submit a proposal for a Tx to be included in one of the blocks of the HLF blockchain. Clients propose a transaction through an application that uses an SDK's (Java, Python etc) API. This is shown as the first step in Figure \ref{fig:hyper}.
   
   \item[2. Endorsement and Tx simulation:] The transaction proposal from the above step is then broadcasted to the endorsing peer nodes in the HLF blockchain network. Each endorsing peer verifies the Tx proposal in terms of its correctness (i.e., its structure, the signatures that it contains, and the membership and permission status of the client that submits the transaction) and uniqueness (i.e., this proposal was not submitted in the past). 
    
    After the above checks comes the \emph{transaction simulation step}. Endorsing peers invoke a relevant chaincode (as specified in the Tx proposal by the submitting client). The execution (as per specific arguments specified in Tx proposal) of this chaincode produces an output against the current state of the database (ledger). Without updating the ledger's state, the output of the Tx simulation is sent back in the form of \emph{proposal response} to the client through the SDK. In Figure \ref{fig:hyper} this is shown by the second step.

    \item[3. Inspection of proposal response:] After the above step the client-side application collects the responses from the endorsement step. Afterwards all the responses are cross checked (in terms of the signatures of the endorsing peers and the content of the responses) to determine if there are any disparities among the content of the responses. If the content of all the responses are the same and according to the pre-defined \emph{endorsement policy} (i.e., number of peers whose endorsements---in terms of their signatures---are necessary) then the client submits this Tx to the Ordering Service (more on it in the next step) that will in turn ultimately update the ledger's state as per the Tx simulation outcome in the last step.
    
    It can also happen that in the Tx proposal, made in the last step, only the current state of the ledger was queried. In this case there will be no need to update a ledger's state and hence there is no submission to the Ordering Service by the client. In Figure \ref{fig:hyper} this is shown by step three.
    
    \item[4. Tx submission to the Ordering Service:] The Ordering Service collects various Txs after the last step via various channels. This is step four in Figure \ref{fig:hyper}. %It then \emph{orders} them according to their receiving times at the Ordering Service. This ordered set of Txs is then included in a block, specific to a channel, which will later be appended in the channel's ledger. Steps four and five in Figure \ref{fig:hyper} shows this process.
    
    \item[5. Tx ordering:] Ordering Service \emph{orders} various Txs according to their receiving times. This ordered set of Txs is then included in a block, specific to a channel, which will later be appended to the channel's ledger. This is covered by step five in Figure \ref{fig:hyper}.
    
    \item[6. Tx validation and committing:]
    In this stage all the peers belonging to a particular channel receive a block containing Txs specific to this channel. Each peer then checks all the Txs in terms of their validity. Valid Txs are those that satisfy an endorsement policy. If the Txs pass the validity test then they are tagged as valid otherwise invalid in a block and then this block is finally appended to the ledger maintained by the peers of this channel. This is covered by step six in Figure \ref{fig:hyper}.
    
    \item[7. Ledger update notification:]
    Finally, after the ledger update in the last step the client of the submitting Tx is notified about the validity or invalidity of the Tx that was included in the latest block of the channel's distributed ledger. This is step seven in Figure \ref{fig:hyper}.
    
    \end{description}

\subsection{Ethereum}
\label{subsec:ethereum}
Ethereum is an open-source blockchain platform that can be used in a public or private setting and adds the provision of building decentralized value-transfer applications (DApps). 
Ethereum builds upon the Bitcoin system and introduces the concept of Ethereum Virtual Machine (EVM). EVM implements the Ethereum protocol (discussed next) which is responsible for handling the state transitions and associated computations 
%(pertaining to a set of smart contracts' state machine) 
without the involvement of third party intermediaries.
The logic that powers a DApp is written in the form of a set of computer programs, so called \emph{smart contracts}, that are being executed by the EVM. The concept of a smart contract can be understood as the algorithmic enforcement of policy agreements among, often mutually non-trusting, peers of a consortium \citep{wood2014ethereum}. A set of smart contracts for a DApp, in turn, can be considered as a state machine, which is executed by the EVM of all the participating nodes. 
While the main Ethereum platform is a public blockchain network, the core platform software is open source and allows developers to configure and deploy a private and permissioned blockchain network (test networks) where only authorized nodes are allowed to participate. 

\subsubsection{Ethereum Protocol}
\label{subsec: eth_protocol}
In Ethereum's ecosystem, there are two main types of entities namely: i) an externally owned account (EOA) with an address and a ii) smart contract written in a contract-specific programming language, such as Solidity, and is compiled into byte code which gets executed by an EVM \footnote{\url{http://ethdocs.org/en/latest/contracts-and-transactions/account-types-gas-and-transactions.html}}. In addition to an EOA, a smart contract is also assigned an address when it is deployed on the blockchain, however, it is used in a nuanced manner when compared to the address usage of an EOA. Anyone in possession of an EOA's address credentials can make a value-transfer transaction with another EOA by specifying its blockchain address. In such transfers the overall systems' state remains unchanged. However, in contrast, it is also possible for an EOA to make a transaction with a smart contract. In these types of transactions a specific function of a smart contract is invoked that usually triggers a state change in the overall EVM. It is also possible that one smart contract invokes a function of another smart contract possibly executing another associated EVM.
It should be noted here that in Ethereum, each time a piece of code is invoked for execution (such as a smart contract's function) all the nodes of the network execute the same piece of code ensuring the correct execution of a program's logic. The state change, in turn, is then recorded in a decentralized manner in the form of mined (more on mining later), which are mutually agreed-upon, blocks ensuring immutability of such records. This way Ethereum enables a trusted and decentralized environment to automate a consortium-based application with trusted value-transfer transactions among the (potentially mutually non-trusting) peers of such a consortium.

Looking closely, a transaction-based state change in Ethereum's ecosystem can be understood with the help of Eq. \ref{eq: state_change} \citep{wood2014ethereum}.

\begin{equation}
\label{eq: state_change}
    \sigma_{t+1} \equiv \Gamma(\sigma_{t}, T)
\end{equation}

In Eq. \ref{eq: state_change}, $\Gamma$ represents a state-transition function and $\sigma$ represents an arbitrary state. A state ($\sigma$), in general, can be defined as a collection of different types of records. As an example a state, in Ethereum, can consist of account balances, operations on a piece of data, specifics of agreements between two transacting parties etc \citep{wood2014ethereum}.

%\textbf{Mining in Ethereum:} Ethereum is using Ethhash\footnote{https://github.com/ethereum/wiki/wiki/Ethash} as a proof-of-work (PoW) algorithm which involves in finding a nonce input to the algorithm so results is a certain difficulty threshhold. The time needed to find a nonce depends on the threshhold. For instance, on average, a block mining takes 15 seconds in Ethereum. Miners (people who do the mining), as a reward for their work are given block rewards (currently 3 ETH per block) and transaction fees (i.e., the gas) are paid by people who wish to have their transcations added to the blockchain. However, because the PoW mining uses a lot of computing power and therefore a lot of electricity, Ethereuem developers came with different consensus models like Proof of Stake (PoS) or Proof of Authority (PoA).

    \textbf{Consensus engines in Ethereum}: Presently Ethereum predominantly uses a \emph{PoW}-based consensus engine called \emph{Ethash}. PoW can be understood as a lottery-based consensus protocol introduced and popularized by Bitcoin \citep{nakamoto2008bitcoin}. The primary purpose of PoW is to avoid double spending of digital assets. PoW provides a trust guarantee to a payee which helps her establish the absence of a double spend of a unit of a digital asset. The actual proof is provided in the form of an integer so called a \emph{nonce} which if, together with all the data contained within a block, is hashed produces an output string of characters which matches a predefined pattern. Such a pattern of hash outputs determines the computational difficulty of finding such a nonce. The process of finding a nonce is referred to as \emph{mining}. More specifically the difficulty of mining a block (i.e., finding a relevant nonce) is determined by number of leading zeros of a hash output. The peer node of a blockchain network who finds a nonce is often referred to as a \emph{miner}.
    
    One of the other consensus engines currently in use in Ethereum's universe is called \emph{Clique}. Clique engine makes use of a consensus protocol called \emph{Proof-of-Authority (PoA)}. In contrast with PoW, PoA is computationally less expensive and eases the process of scaling a network. PoA-based consensus engines help to establish a private and permissioned version of a blockchain. In PoA, in contrast with PoW, the nodes who can have a say in appending new blocks to a blockchain are carefully chosen with known identities are referred to as \emph{sealers}. In turn, the process of appending a new block to a blockchain running a PoA-based consensus engine is called \emph{sealing}. Such nodes are also sometimes referred to as \emph{authorities}. Specifically, sealing implies that if a block contains digital signatures of majority of authorities then it is considered as a valid block. However, PoA has proven to be less secure as compared to PoW\footnote{\url{https://medium.com/poa-network/exploiting-consensus-vulnerability-in-a-parity-client-to-hard-fork-an-ethereum-based-network-d2c368bf0bac}} and that is the reason that it is predominantly being used by test networks and private chain setups for experimental purposes. In this paper we also make use of PoA-based Clique engine for our experiments. Since we have setup a local and private Ethereum blockchain we conjecture that security is not going to be an issue as far as our empirical analyses are concerned.

    As a final note on Ethereum's consensus engines, we would like to briefly mention \emph{Proof-of-Stake (PoS)}. Ethereum's community is planning an eventual migration from its PoW-based Ethash to a PoS-based consensus engine primarily because of network scalability issues prevalent in a computationally intensive PoW. PoS can be understood as close to PoA where instead of an identity of a node the monetary value in the form of digital assets that a node owns in the network is at stake. The nodes with biggest stake in the network will have a bigger say when it comes to appending a new block to the chain, while in PoA validators just take turns. However, the idea of a PoS-based Ethereum is still in its infancy and does come with its fair share of problems most notably a mismatch of interests of nodes in the underlying network with equal stake in the network\footnote{\url{https://medium.com/poa-network/proof-of-authority-consensus-model-with-identity-at-stake-d5bd15463256}}.

%\end{description}

%\begin{figure}[h]
%\centering
%\includegraphics[width=6.5in,keepaspectratio]{images/ethereum_pr.pdf}
%\captionsetup{justification=centering}
%\caption{Ethereum Protocol}
%\label{fig:ether}
%\end{figure}

\subsection{Comparison of Hyperledger Fabric and Ethereum}
\label{subsec: hlfVSeth}
%
 %\mennan{We need to show some parameter !}
%We believe that t
The main differences between HLF and Ethereum derive from their inherent approaches of adopting a permissioned vs open blockchain paradigm respectively. HLF, as we see above in Section \ref{subsec: hlf}, has been developed mainly to promote a closed and permissioned version of blockchain with stringent confidentiality guarantees among a set of transacting member peers, by segregating them in different channels and associated chain code or a distributed application. On the other hand Ethereum proposes an open and generic flavour of blockchain, as a platform to build distributed applications and promote automation. The evolution of both of these platforms is heavily influenced by the original premises described above.

%\begin{description}

\textbf{Consensus}: In its first incarnation, i.e., in Bitcoin \citep{nakamoto2008bitcoin}, Proof of Work (PoW) based consensus mechanism was originally proposed to solve the double-spending problem (see Section \ref{subsec:ethereum} for details) in an asynchronous manner. Tracing the evolution ladder up one step, Ethereum can be considered as an evolved and more generic version of Bitcoin which popularized the concept of smart contracts with an associated Turing complete programming language. Since Ethereum builds upon Bitcoin's core, it is quite natural for it to inherit many of the Bitcoin's legacy traits: PoW-based asynchorous consensus mechanism, being public and permissionless are among the most notable ones. We did, however, see Ethereum making use of a Proof of Authority (PoA) based consensus engine in the last section to introduce a permissioned flavour of its blockchain. Such proposals are still in their infancy and we conjecture that they need time to get mature and widely adopted. In comparison, HLF has adopted a more hierarchical and synchronous approach in achieving consensus and appending new blocks to a blockchain. This is mainly due to the Ordering Service (as described in Section \ref{subsec: hlf}), which is responsible to timestamp the transactions and avoid double-spends in the network. The same Ordering Service introduces, up to some extent, centralization in the way HLF achieves consensus on a particular state of a blockchain. If a single instance of an Ordering Service is used then it raises the concern for a single-point-of-failure as well. However, single-point-of-failure is not a major concern in the asynchronous PoW-based blockchains such as Ethereum. In our experiments we deploy only one instance of Ordering Service to keep the setup simple and perform experiments to evaluate HLF's core architecture rather than auxiliary problems like the single-point-of-failure.

\textbf{Architecture}: There are several aspects to consider: 
\begin{itemize}
\item{\textit{Public vs private}:} HLF and Ethereum's design differ in their approach in addressing the pool of usecases in public and enterprise domain. Ethereum was designed to be completely decentralised setup in public domain with all nodes in the network being equal and then slowly being adopted to private or enterprise usecases. Ethereum also has a native currency called \textit{ether}. HLF was designed focusing to solve enterprise usecases, rather than providing a public blockchain platform. It does not have inbuilt currency, but allows for creation of coins on top of the network through chaincode.
\item{\textit{Extent of Decentralisation}:}
Not all nodes are equal in HLF's network. There are set of privileged nodes - \textit{endorsers, orderer, membership providing service}, that have more control and access than other nodes - \textit{peers}, making the system more centralised as compared to Ethereum.  Further, in contrast with Ethereum's flat approach to reaching consensus, HLF has two main interlinked levels where consensus is reached in an hierarchical order. HLF follows an order of \textit{execution of the chaincode  and validation of transaction with endorsement policy by multiple endorsers, order the transaction and put in a block}. More specifically, the first level involves satisfying  an endorsement policy where signatures from a pre-set number of endorsing peers are collected for a transaction proposal (see Section \ref{subsec: hlfProtocol} for details). The second step occurs at the level of Ordering Service, which orders the transactions in block.  On the other hand, Ethereum follows an order where \textit{ each node should check and execute the transaction, generate/propose a block, validate the block and broadcast it}. The initial check happens by the node where the transaction is submitted, and not by a set of endorsing nodes as in case of Hyperledger Fabric. This approach of Ethereum leads to higher collision during reaching consensus, as well as to higher chances of forks in a large network. Ethereum handles forking using GHOST protocol. While HLF's design makes it less prone to forking and with careful design of endorsement policies, the forking problem can be completely avoided. At the ordering level, HLF provides a modular plug-and-play approach where a consensus mechanism can be chosen from a set of available mechanisms that can be deployed pertaining to a specific use case at hand. Currently the default option is a single orderer setup, while Kafka based multi-orderer setup and PBFT techniques are also popular.
\item{\textit{Confidentiality}:}
 When it comes to the permissioned blockchain paradigm, HLF's approach to implement a private and permissioned blockchain is more exhaustive and fine grained as compared to Ethereum. As we discussed in Section \ref{subsec: hlf}, channels and MSPs implement an intricate and hierarchical, much akin to an actual organization, permissioned infrastructure with clearly defined roles and security clearance for different entities in the network. In Ethereum, the closest deployment to a permissioned blockchain would be adopting a PoA-based consensus engine, which is still quite a flat approach as compared to HLF's MSPs and channels with their dedicated ledgers at different levels with associated set of chain codes.

\end{itemize}
\section{Case study: A local blockchain in the QMPSU mesh network}
\label{sec:qmp}

The Quick Mesh Project (qMp) \footnote{\url{http://qmp.cat/Overview}} develops a firmware based on OpenWrt Linux with the aim to ease the deployment of mesh networks by the users who are willing to interconnect in an area, and pool their Internet uplinks \citep{LlorencMSWiM}. qMp was initiated in $2011$ by a few Guifi.net activists. 

The qMp firmware has enabled to deploy several mesh networks with actual end-users (e.g., more than 250 active locations, typically households) in several parts surrounding the city of Barcelona\footnote{\url{http://dsg.ac.upc.edu/qmpsu}}. At the time of this writing, there are $10$ different neighbourhood mesh networks, and the largest (Sants-UPC or QMPSU) has $85$ operational nodes. In that network, there are two gateways that connect the QMPSU network to the rest of guifi.net and the Internet. Users join the mesh by setting up \emph{outdoor routers} (i.e., antennas) that automatically establish router-to-router links. The outdoor routers are connected through Ethernet to a home network, with an indoor AP (access point) where the edge server devices run diverse services: home-servers such as Raspberry Pi's or Cloudy devices \citep{Baig:2018}.
The network and the home-servers are deployed and operated by citizens that coordinate in periodic meetings and a mailing list. These home-servers are used for network management and traffic accounting, and the fact of having several independent traffic logs and network monitors, even using different software tools, which allows to increase resilience but also trust, as not all can fail, lose data due to failures or network partitions, or collude at the same time.

\begin{figure}[!h]
    \centering
    \begin{minipage}{0.47\textwidth}
        \centering
        \includegraphics[width=3.1in,keepaspectratio]{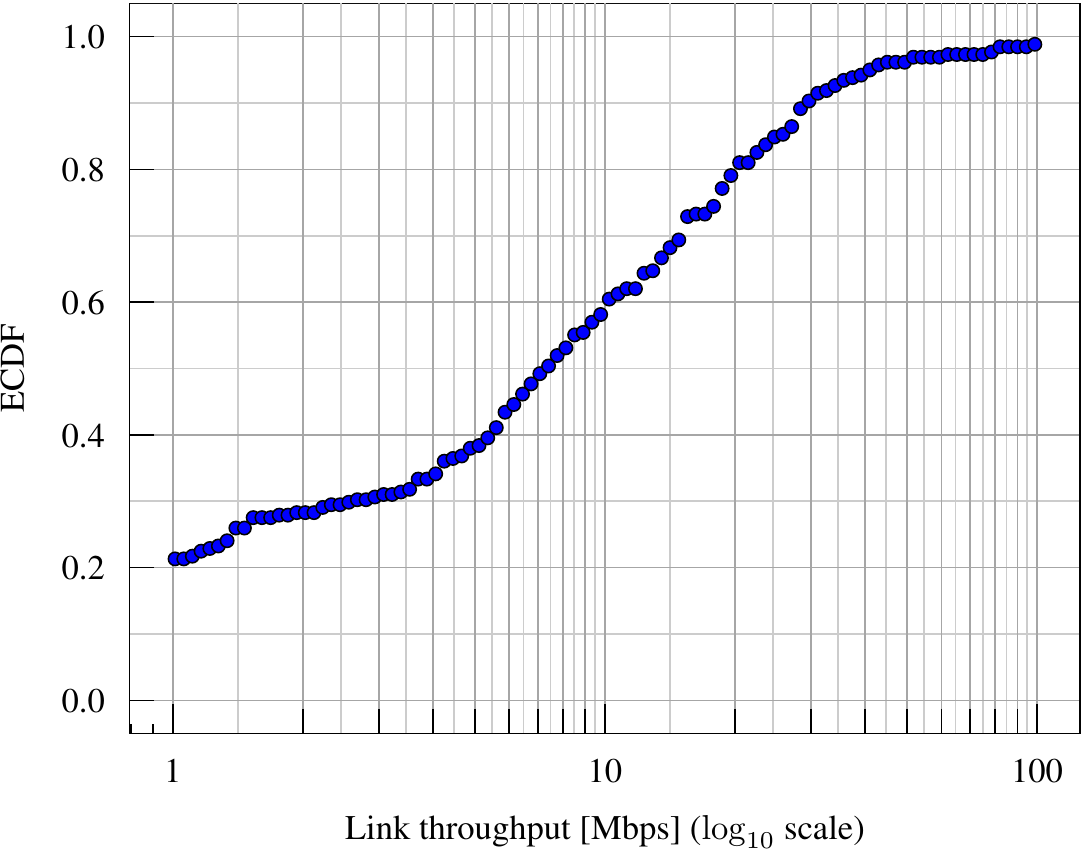}
        %\vspace{-.2cm}
        \caption{Bandwidth ECDF}
        \label{fig:bwd}
    \end{minipage}%
    \begin{minipage}{0.47\textwidth}
        \centering
        \includegraphics[width=3.1in,keepaspectratio]{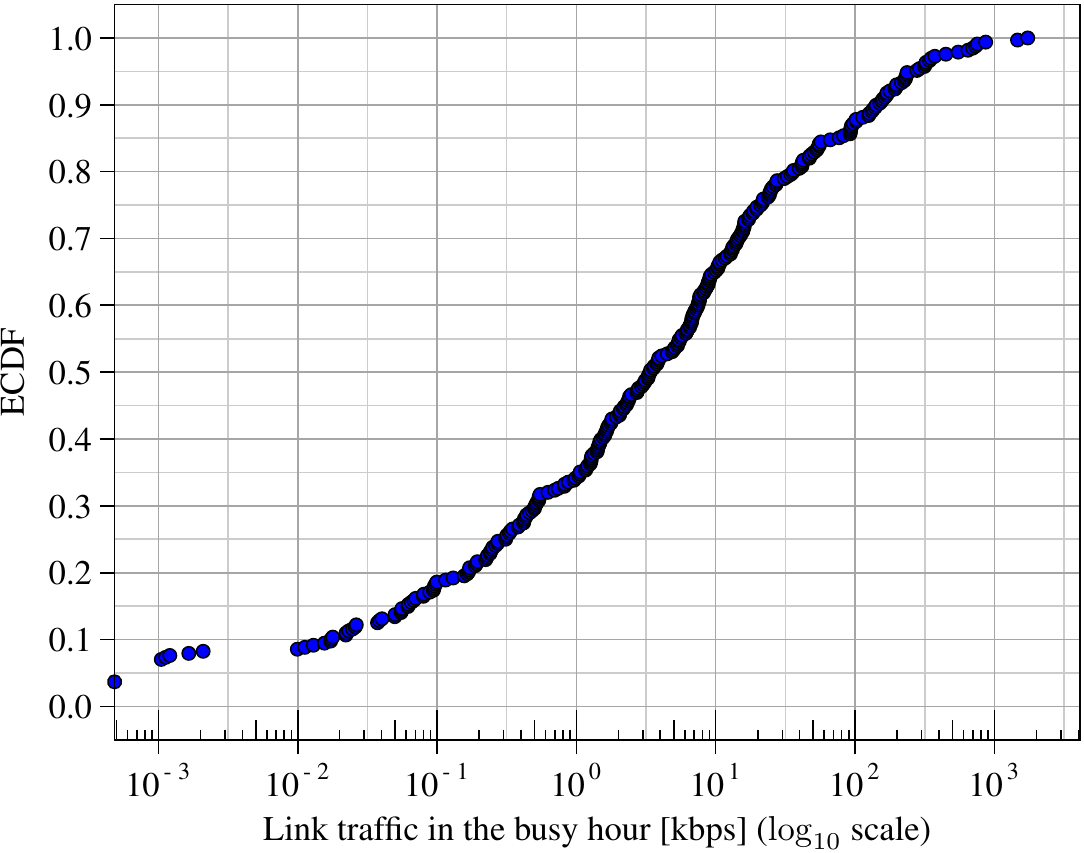}
        %\vspace{-.2cm}
        \caption{Traffic ECDF}
        \label{fig:tfd}
    \end{minipage}
    %\caption{Bandwidth and traffic distribution in the qMp}
    \label{fig:bwtf}
\end{figure}

%%%%%%%%%%%%%
%%%%%%%%%%%%%
%%%%%%%%%%%%%
%\begin{figure}
   % \centering
%    \begin{subfigure}[!]{0.2\textwidth}
%        \includegraphics[width=1.6in,height=1.2in]{img/bw_march.pdf}
%        \caption{Link throughput [Mbps] ($log_{10}$ scale)}
%        \label{fig:bwd_l}
%    \end{subfigure}
    %~ %add desired spacing between images, e. g. ~, \quad, \qquad, \hfill etc. 
    %\begin{subfigure}[!]{0.2\textwidth}
     %   \includegraphics[width=1.6in,height=1.2in]{img/tf_march.pdf}
      %  \caption{Link traffic in the busy hour [kbps] ($log_{10}$ scale)}
       % \label{fig:bwd_r}
    %\end{subfigure}
%    ~ %add desired spacing between images, e. g. ~, \quad, \qquad, \hfill etc. 
 %   \caption{Bandwidth and traffic distribution in the QMPSU}\label{fig:bwtf_t}
%\end{figure}

\textbf{Network performance:} We monitored the QMPSU mesh network for a period of one month. We took hourly captures from the network for the entire month of March $2018$. Figures \ref{fig:bwd} and \ref{fig:tfd} depict the bandwidth and traffic distribution of all the links in the network. Figure \ref{fig:bwd} shows that the link throughput can be fitted with a mean of $13.6$ Mbps. At the same time Figure \ref{fig:bwd} reveals that $60\%$ of the nodes have $10$ Mbps or less throughput. Figure \ref{fig:tfd} demonstrates that the maximum per-link traffic in the busiest hour is $1736$ kbps. We observed that the resources are not uniformly distributed in the network. There is a highly skewed bandwidth and traffic distribution.                      
 
\begin{figure}[!h]
    \centering
    \begin{minipage}{0.5\textwidth}
        \centering
        \includegraphics[width=.85\linewidth]{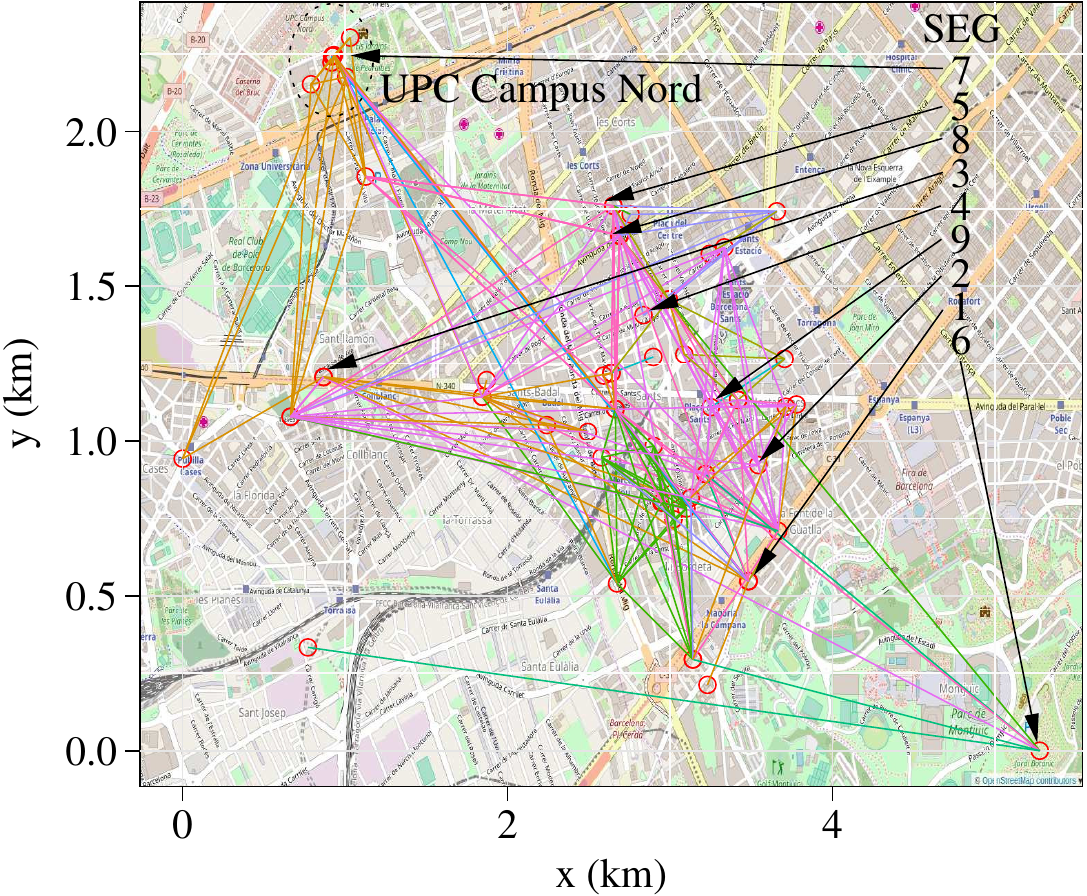}
        \vspace{0.3cm}
        \caption{Topology of the deployed nodes in Barcelona.}
        \label{fig:topo}
    \end{minipage}%
    \begin{minipage}{0.5\textwidth}
        \centering
        \includegraphics[width=0.85\linewidth]{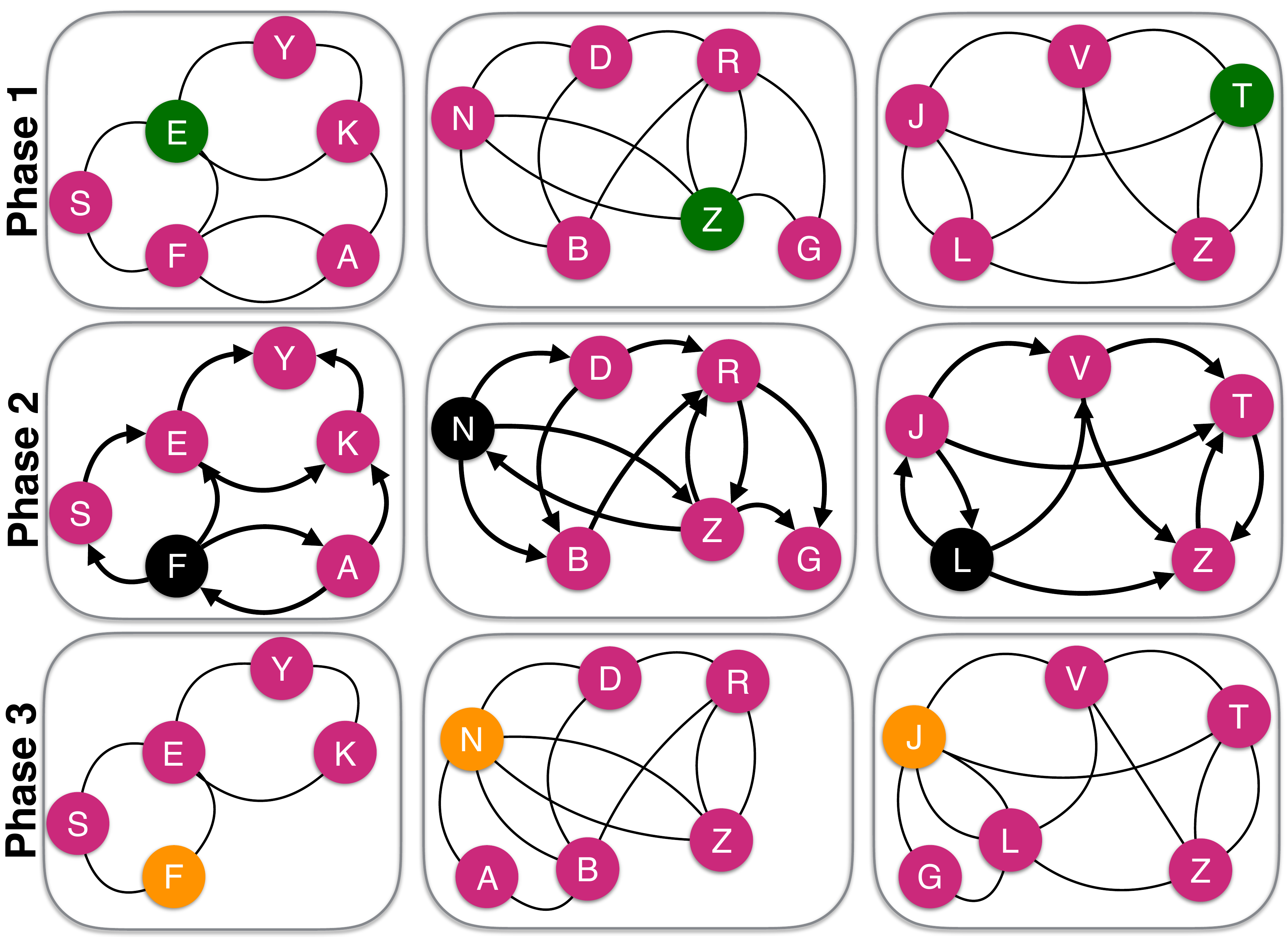}
        \vspace{0.9cm}
        \caption{BASP Phases}
        \label{fig:basp}
    \end{minipage}
    %\caption{Sealing Time \textbf{(a)} and Completion Time \textbf{(b)} for 100 Transaction.}
    %\label{fig:bwtf}
    %\caption{Mennansas as}
\end{figure}

\textbf{Node deployment:} Based on the network measurement analysis we strategically deployed $10$ Raspberry Pi (RPi3)  devices on the outdoor routers to cover the area of the QMPSU network as presented in Figure \ref{fig:topo}. We use our previous work \citep{Selimi2018} on service placement to determine nodes in the network. In this set, we cover nodes with different properties: with higher bandwidth \citep{Selimi2018}, nodes that are highly connected (i.e., with high degree centrality) \citep{Gelly2018}, nodes acting as bridges (with high betweenness centrality), and nodes not well connected. After the nodes were chosen, we deployed $10$ RPi boards in the community users home.

%\begin{figure}[tbh]
%\centering
%\includegraphics[width=4in,keepaspectratio]{qmpRPI.pdf}
%\captionsetup{justification=centering}
%\caption{Topology of the deployed Raspberry Pi nodes in Barcelona.}
%\label{fig:topo}
%\end{figure}

\textbf{BASP: Bandwidth Aware Service Placement}
In order to determine the best nodes in the QMPSU network where to place the Hyperledger Fabric and Ethereum components, we use the BASP heuristic from our previous work \citep{Selimi2018} and \cite{SelimiCCGrid}. The BASP (Bandwidth and Availability-aware Service Placement) service placement heuristic takes into account the bandwidth of the network, node availability and CPU of the nodes to do smart node selection/placement. BASP is executed every single time a (new) service or node deployment is about to be made. BASP runs in three phases. In the first phase, BASP partitions the network topology into $k$ (maximum allowed number of service replicas) and removes the nodes that are under the pre-defined availability threshold. In this phase, BASP uses the naive K-Means partitioning algorithm in order to group nodes based on their geo-location. The idea is to get back clusters of nodes that are close to each other. In the second phase, BASP estimates and computes the max bandwidth of the nodes in the network. The bandwidth between two nodes is estimated as the bandwidth of the link having the minimum bandwidth in the shortest path. In the third phase, BASP re-assigns nodes with higher CPU and availability to the selected clusters formed in the second phase. Figure \ref{fig:basp} demonstrates the phases of the BASP.

%\begin{figure}[tbh]
%\centering
%\includegraphics[width=4.1in,keepaspectratio]{crytopo.pdf}
%\captionsetup{justification=centering}
%\caption{Topology of the deployed nodes}
%\label{fig:topo}
%\end{figure}

%\begin{figure}[tbh]
%\centering
%\includegraphics[width=2.8in,keepaspectratio]{img/lqdc.pdf}
%\captionsetup{justification=centering}
%\caption{Link Quality vs Degree Centrality}
%\label{fig:lqdc}
%\end{figure}

\section{Evaluation}
\label{sec:eval}

We evaluate the performance of Hyperledger Fabric and Ethereum PoA inside a wireless mesh setting for parameters like CPU load, memory consumption and transaction latency with varying number of transactions as well as varied placement strategies of blockchain nodes. For this, we setup a testbed network comprising RPi3 boards in the QMPSU network\footnote{http://dsg.ac.upc.edu/qmpsu/index.php}. Each RPi3 board runs either a Ethereum node or a component of Hyperledger Fabric (see Section \ref{subsec: hlf} and Section \ref{subsec:ethereum} for details). The RPi3 boards have 1.2GHz 4 core ARM cortex A53 processor, a RAM memory of 1GB and run \emph{raspbian-stretch} OS. In parallel, we also deployed a similar setup in the lab environment (for performance comparison purposes) and evaluated the performance in both environments. 

We perform identical experiments in permissioned blockchain network setup for both Ethereum PoA and Hyperledger Fabric. The scenario of a typical experiment in both Ethereum PoA as well as Hyperledger Fabric is - a client sends \emph{N} balance transfer request between two parties/accounts to a node in the blockchain network. In Ethereum \textit{sendTransaction} operation is used to transfer funds between two accounts. While in the case of Hyperledger Fabric, a chaincode, deployed in endorsers, is executed to transfer funds between two parties e.g., Alice sends 10 tokens to Bob [A, B, 10].

%For Ethereum, Proof of Authority consensus mechanism \textbf{Clique}, implemented in geth is used and experiments are performed with $1$, $2$ and $4$ sealers/validators. In case of Hyperledger Fabric, single orderer for consensus and ordering the transactions in the blockchain. Experiments are performed with $1$, $2$ and $4$ endorsers. 

% In both Ethereum and Hyperledger Fabric, $100$ such requests are fired in parallel creating a load on the blockchain node and blockchain network. %Measurements of CPU load, memory usage and transaction latency are noted. Also, transaction latency measurements are made by varying number of endorsers (in case of Hyperledger Fabric)/ sealers (in case of Ethereum) with nodes placed randomly and with nodes placed in best position as determined by the BASP heuristic. In case of Ethereum we also measure transaction latency for varying number of requests N = {$1$, $10$, $100$, $1000$}.

\subsection{Evaluation Metrics and Transaction Application}
In order to evaluate the targeted blockchain platforms, we have designed a simple money transfer application where money/tokens are exchanged between two parties in the blockchain network. The money transfer application is deployed in the blockchain network. In case of Hyperledger Fabric, the application is deployed as a chaincode, while in Ethereum, we run the application as a $nodejs$ container that connects to Ethereum network using web3 framework to make transactions. In both the platforms, as part of initialisation, we create user accounts - among who the money/token transfer takes place, setup balances in the accounts and make the setup ready for transactions. In Ethereum, we use the existing methods to create accounts and setup balances. In Hyperledger, the deployed chaincode has methods to create accounts, setup balances along with method to transfer money. The function to transfer money is called $sendMoney$. In Ethereum we use the existing $sendTransaction$ method to transfer money between accounts while a custom sendMoney function, as part of chaincode, is called to transfer balances in Hyperledger Fabric. The following code snippets (Transaction code 1 and 2) detail the custom $sendMoney$ function deployed in Hyperledger Fabric and $sendTransaction$ method in Ethereum. For chaincode in Hyperledger Fabric, we have used the example chaincode given by Hyperledger group, modified it according to our requirements. On top of it, scripts are written, which are executed from a client node in Hyperledger Fabric network to make transactions repeatedly. Below, we present the code snippets for 
\begin{enumerate}
    \item Hyperledger Fabric: the $sendMoney$ function, which is part of chaincode\footnote{\url{https://github.com/anirudhkabi/HLF/blob/master/chaincode/high-throughput.go}} and is used to make transactions, and a script that is used for repeated transactions. The procedure is invoked\footnote{Invocation: peer chaincode invoke -o orderer.example.com:7050 -C \$CHANNEL\_NAME -n  \$CHAINCODE\_NAME -c \{Args:[update,\$1,\$2,\$3]\}} with appropriate input parameters.
    \item Ethereum: A $nodejs$ procedure\footnote{\url{https://github.com/DSG-UPC/EthereumMeasurements}} that is called to invoke multiple times the inbuilt token transfer method $sendTransaction$.
\end{enumerate} %We implement each function separately in the platforms. During pre-configuration, user accounts are created and smart contracts are deployed in each platform ready to be invoked during evaluation period. Smart Contract 1 and 2 depicts a code snippet from SendMoney and SendTransaction functions.  

\begin{algorithm}[b]
  \caption{Code snippet to invoke the SendMoney function multiple times in Hyperledger Fabric}\label{hypercode}
  \begin{algorithmic}[1]
    \Procedure{sendMoney}{$Name, value, op$} \Comment{Code snippet written in Go lang as part of Chaincode} %\label{hypercode1}
        \State $txid := APIstub.GetTxID()$;
        \State $compositeIndexName := NameOpValueTxID$;
        \State $CompositeKey = APIstub.CreateCompositeKey(compositeIndexName, []string{name, op, Value, txid})$
        \State $APIstub.PutState(compositeKey, []byte{0x00})$
    \EndProcedure
   \Procedure{manyUpdates}{$Name,value,op,repetitions$} \Comment{Script to Invoke the SendMoney for multiple transfers}
    \For{{\texttt{let $i = 0$; $i<repetitions$; $i++$}}}
        %\State $do$
        \State \Call{sendMoney}{$Name, value, op$} \Comment{Chaincode invocation}
    \EndFor
   \EndProcedure
  \end{algorithmic}
\end{algorithm}

\iffalse
\begin{algorithm}
  \caption{Code snippet for SendMoney function, written in Go lang as part of Chaincode in Hyperledger Fabric}\label{hypercode}
  \begin{algorithmic}[1]
    \Procedure{sendMoney}{$Name, value, op$}%
        \State $txid := APIstub.GetTxID()$;
        \State $compositeIndexName := NameOpValueTxID$;
        \State $CompositeKey = APIstub.CreateCompositeKey(compositeIndexName, []string{name, op, Value, txid})$
        \State $APIstub.PutState(compositeKey, []byte{0x00})$
         
    \EndProcedure
  \end{algorithmic}
\end{algorithm}

\begin{algorithm}
  \caption{Script to Invoke the SendMoney for multiple transfers}
  \begin{algorithmic}[1]
   \Procedure{manyUpdates}{$Name,value,op,repetitions$}
    \For{{\texttt{let $i = 0$; $i<repetitions$; $i++$}}}
        %\State $do$
        \State peer chaincode invoke -o orderer.example.com:7050 -C \$CHANNEL\_NAME -n  \$CHAINCODE\_NAME -c {Args:[update,\$1,\$2,\$3]}
    \EndFor
   \EndProcedure
  \end{algorithmic}
\end{algorithm}
\fi 

%\newpage
%solved, sorry. I dont think so it solves the issue

\begin{algorithm}
  \caption{Snippet for nodejs script calling SendTransaction function in Ethereum}\label{ethercode}
  \begin{algorithmic}[1]
    \Procedure{sendTransaction}{$nonce, repetitions, etherValue, initSecond$}%\Comment{The g.c.d. of a and b}
      \For{{\texttt{let $id = 0$; $id<repetitions$; $id++$}}}
         \State $nonce = nonce+1$;
         \State $times[id] = {} $;
         \State $times[id]['id'] = id$;
         \State $times[id]['start'] = Date.now()$;
         \State \Call{web3.eth.sendTransaction}{from: account, to: acc2, value: web3.utils.toWei(String(1)),
      gas: 3000000)}
          .on('confirmation', function(confirmationNumber, receipt)
            \If{$confirmationNumber==0$}
      \State $times[id]['mindTime']=Date.now()-times[id]['start']$;
          \State $console.log(times[id])$;
      \EndIf
      \If{$confirmationNumber==12$}
      \State $times[id]['completionTime']=Date.now()-times[id]['start']$;
        \State $console.log('ID: '+id)$;
        \State $console.log(times[id])$
      \EndIf
      \EndFor
      \State \textbf{return} $TransactionReceipt$%\Comment{The gcd is b}
    \EndProcedure
  \end{algorithmic}
\end{algorithm}
 
We consider the following parameters to evaluate the targeted blockchain platforms:
\begin{enumerate}
    \item \textbf{Transaction Latency:} This is the time taken to complete a set number say N = $100$ transactions. Transaction latency is measured differently in Hyperledger Fabric and Ethereum PoA platforms. %\mennan{Sealing time and completion time not explained}
    \begin{itemize}
        \item In Hyperledger Fabric, this is the time taken to endorse and to commit a transaction to the ledger. We plot both time to endorse and time to commit in our plots.
        \item In Ethereum PoA, this is the total amount of time to execute, seal and confirm the transaction. We consider a  total of $12$ confirmations as  successful commit of a transaction \citep{weber2017availability}.
    \end{itemize}
    
    \item \textbf{CPU and Memory Utilization:} Load on CPU and memory is measured for nodes when idle and during transactions.
    
    \begin{itemize}
        \item In Hyperledger Fabric, CPU load and memory usage are measured for endorser, orderer and committing peers. 
        \item In Ethereum PoA, CPU load and memory usage are measured for the sealer node and non sealer node to which the transaction is fired to. 
        
    \end{itemize}

    \end{enumerate}

\subsection{Hyperledger Fabric Setup }

In our experiments, we deploy a HLF blockchain network\footnote{\url{https://github.com/anirudhkabi/HLF}} consisting of a single organizational entity. All the transactions happen among the members of this single organization. The HLF components, namely peer (we deploy multiple instances of this component), orderer, and client are deployed in different RPi3 boards connected to each other in QMPSU wireless mesh network. We perform experiments by placing different Hyperledger Fabric components at different physical (RPi3) nodes and by varying the number of peers from $1$ to $4$. We evaluate the setup in QMPSU network  comparing transaction latency for $100$ transactions fired in parallel when components of HLF are placed randomly in the network and in nodes with best connectivity (according to BASP). For the best connectivity deployment, we also evaluate transaction latencies in HLF for a $2$ peer setup when the block size is varied from $10$ to $100$ transactions per block. Our experiments comprise of $5$ runs (taken in different time slots) and the presented results are averaged over all the runs.

\subsubsection{Hyperledger Fabric Experimental Results}

%\subsubsection{Transaction latency}
%\label{subsubsec: txLat}
%
Table \ref{table:table1} lists the transaction completion time (referred to as \emph{Time-to-Commit (TCC)}) for $100$ transactions, initiated in parallel, between the two peer nodes in the lab environment and in the QMPSU network respectively with block sizes ranging from $10$ to $100$ transactions per block. It can be observed that, as the block size increases, the transaction completion time increases in the QMPSU network.

\begin{table}[ht]
\centering
\begin{tabular}{c | c | c | c p{5cm}}
\hline\hline
 \textbf{Block Size}&\textbf{Time-to-Commit (Lab)}& \textbf{Time-to-Commit (QMPSU)} & \textbf{\# of Txs}    \\ 
\hline 
     10     & 33.4 s      & 64.2 s & 100  \\
     20     & 35.0 s    & 69.7 s & 100     \\
     50     & 39.2 s     & 75.3 s & 100    \\
     100 & 45.3 s & 84.8 s & 100 \\
 
 \hline 
\end{tabular}
\caption{Transaction delivery time (parallel transactions).} 
\label{table:table1}
\end{table}

%\begin{table}[ht]
%\centering
%\begin{tabular}{c | c | c | c p{5cm}}
%\hline\hline
% \textbf{Block Size}&\textbf{TTC(Lab)}& \textbf{TTC(QMPSU)} & \textbf{\# of Tx}    \\ 
%\hline 
%     10     & 48.88 sec      & 89.39 sec & 100  \\
%     20     & 50.30 sec    & 92.05 sec & 100     \\
%     25		& 51.12 sec   & 93.60 sec  & 100  \\
%     50     & 56.58 sec     & 105.82 sec & 100    \\
%     100 & 66.29 sec & 121.29 sec & 100 \\
 
\begin{figure}[t]
\centering
\includegraphics[width=3.5in,height=2.3in]{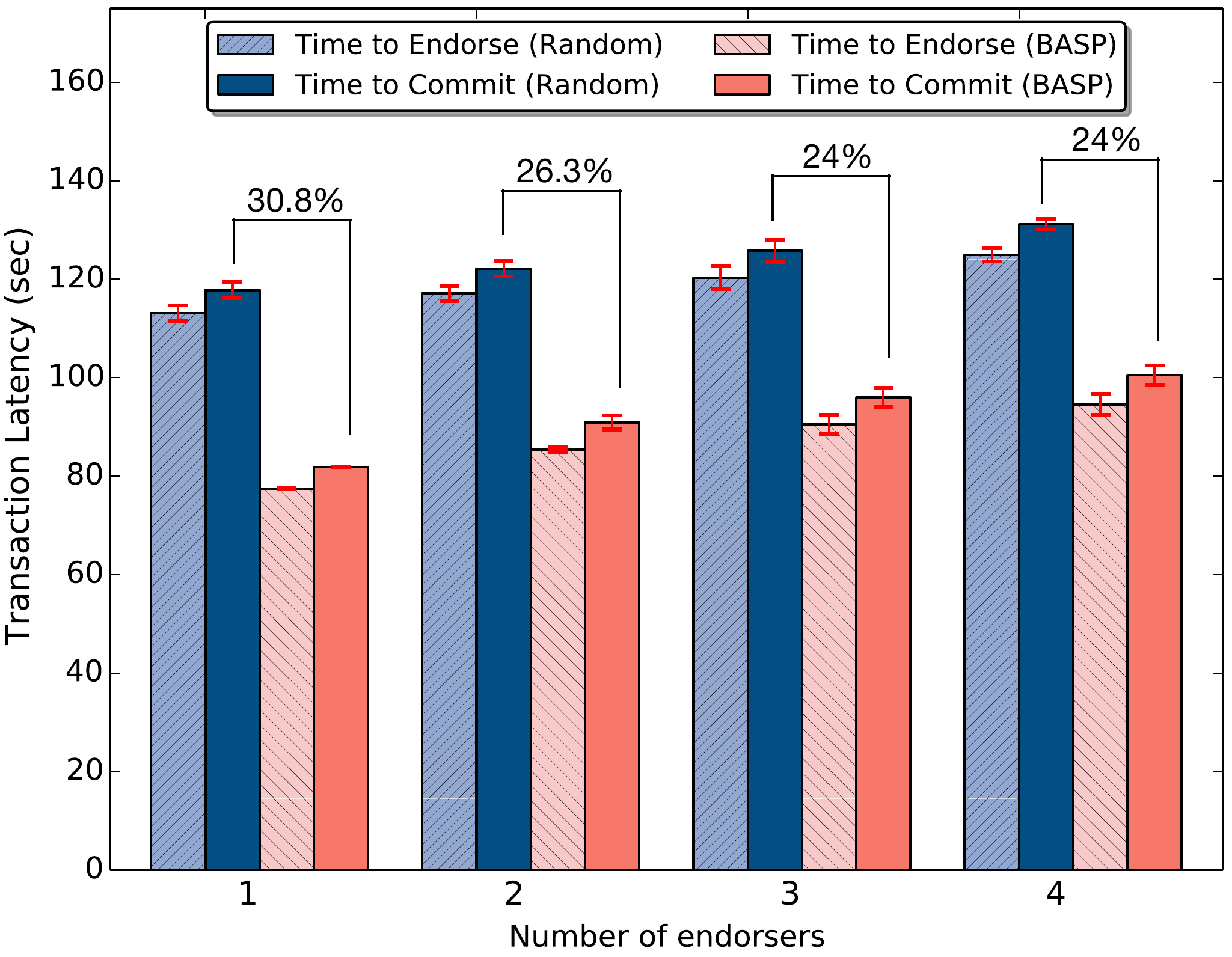}
\captionsetup{justification=centering}
\caption{Transaction latency (QMPSU, HLF)}
\label{fig:comp_qmp}
\end{figure}

% \hline 
%\end{tabular}
%\caption{Transaction delivery time (Lab and QMPSU setup)} 
%\label{table:table1}
%\end{table}
%
\textbf{Transaction Latency}: In Hyperledger Fabric, transaction latency is defined as the total time taken to endorse and to commit a transaction to the ledger. Figure \ref{fig:comp_qmp} shows the comparison of transaction latency observed for two different placements of HLF ordering service. We measure transaction latency when the HLF ordering service is placed randomly in the network (Random) and when it is placed at the node chosen with a heuristic that considers the node with higher bandwidth and degree centrality (BASP) \citep{Selimi2018}. The results of Figure \ref{fig:comp_qmp} are obtained when a client initiates $100$ transactions sequentially. This Figure reveals that the gain brought by BASP, for the case when we have one endorser in the network, is a  $30.8\%$ reduction. For the case when we have four endorsers in the network, the gain of BASP over Random is $24\%$ reduction. 
%\lnm{not sure how that below can be seen. I leave a milder sentece instead}
%Further, Figure \ref{fig:comp_qmp} demonstrates that in the QMPSU network it takes $1.2$ seconds for a single transaction to be appended to the distributed ledger. 
Further, we can deduct from Figure \ref{fig:comp_qmp} that in the QMPSU network it takes around $1$ second for a single transaction to be appended to the distributed ledger. 

%{\color{red}(it would be good if we could also provide more information about the error bars in Figure \ref{fig:comp_qmp}. As in what confidence interval was used?)}.

\begin{figure}[tb]
\centering
\includegraphics[width=5in,keepaspectratio]{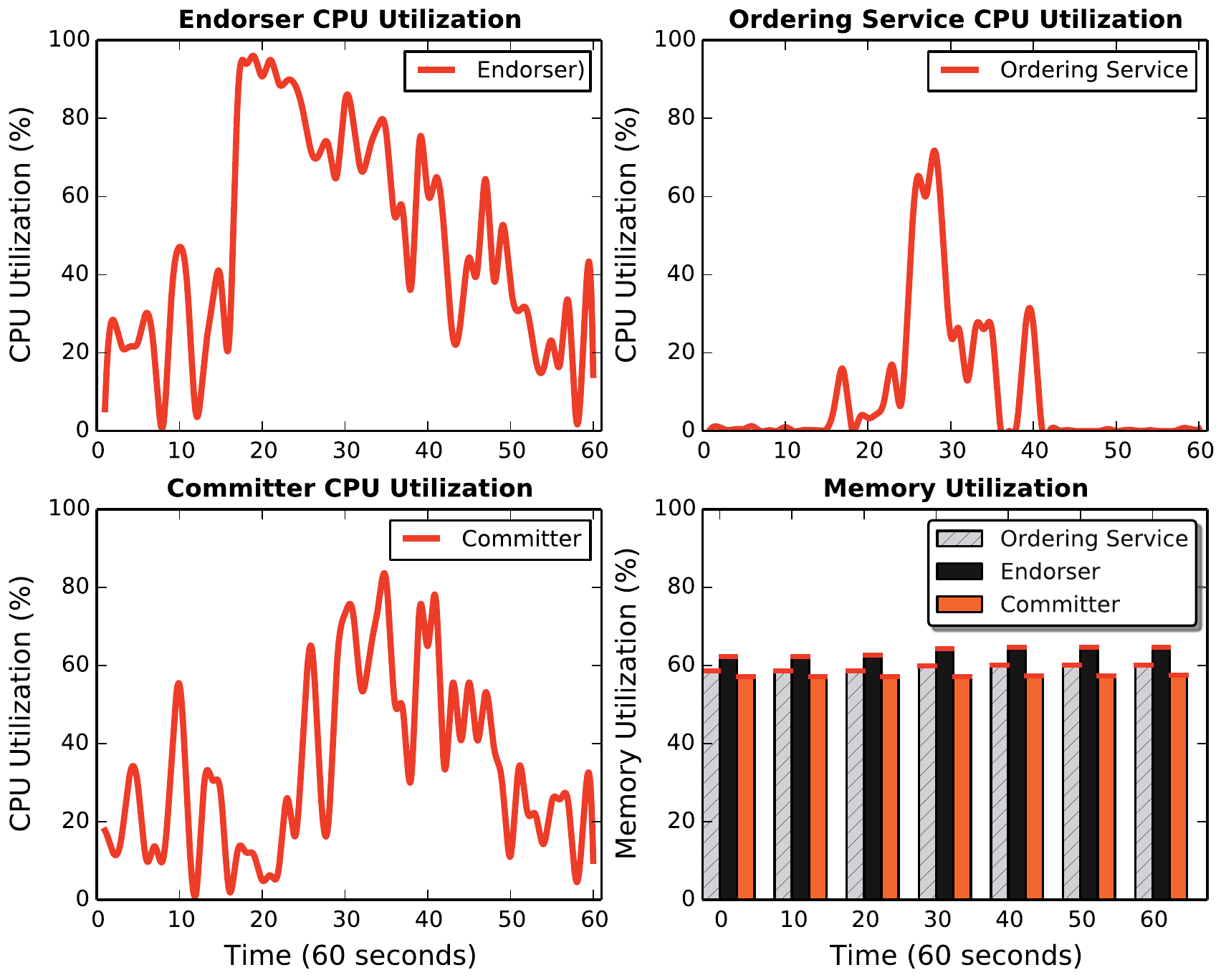}
\captionsetup{justification=centering}
\caption{CPU and memory utilization (HLF)}
\label{fig:cpu_memory}
\end{figure}

%\subsubsection{Resource consumption}
%\label{subsubsec: res_cons}
%
\textbf{Resource Consumption:} Figure \ref{fig:cpu_memory} shows CPU utilization by various components of the HLF network namely: an orderer, a client and two peers (an endorser and a committer). CPU utilization of all nodes is monitored for a time period of $60$ seconds during which $100$ transactions are fired in parallel (by the client) and all the transactions are completed. $100$ parallel transactions took around $40$ seconds to complete. We chose to monitor the nodes for a time period of $60$ seconds to show idle phase usage and busy phase usage of each node. In the graph, transactions are initiated at the $11$th second and all the transactions get completed at $50$th second. %{\color{red}(do we know that all 100 of transactions get completed within the window of 60 seconds?)}
It can be observed that the endorser is the node with the highest CPU utilization whereas the orderer utilizes the least of CPU. %{\color{red}(Suggestion: In the these three graphs can we also provide an average value of CPU utilization as well?)}. 
Figure \ref{fig:cpu_memory} shows that, for $100$ transactions initiated at the same time, the endorser's maximum CPU utilization reaches $96\%$. The maximum CPU utilization is $81\%$ for the committer while it is $71\%$ for the orderer. The reason that the endorser has the highest CPU consumption, among other HLF components, is because of the chaincode execution at the endorsing peer, which does not happen at the committer and the orderer. \\

In HLF, each component usually runs in it own Docker container\footnote{\url{https://www.docker.com/what-docker}}. The chaincode container executes the chaincode for each incoming transaction which is something that does not happen at the committer node. When multiple transactions take place in parallel, concurrent execution of the chaincode happens for all transactions thus, in turn, increasing the load on the endorsing peer. With $100$ parallel transactions, we observe that the CPU load reaches $96\%$ at the endorser. However, the load on each endorser can be reduced by deploying multiple endorsers in the network. The load on different endorsers can be balanced by designing a suitable endorsement policy and devising a strategy at the client to request endorsements from different set of endorsers each time a transaction is initiated.

Similarly, memory usage is the highest by the endorser and the least by the orderer. Memory usage of committing peer falls in between of endorsing peer and the orderer. At the orderer and the committing peers, memory usage remains almost the same level between the idle phase and during transaction execution. Memory usage at the orderer mostly falls in the range of $57\%$-$58\%$ while the memory usage at the committer is in the range of $57\%$-$60\%$. At an endorsing peer the memory usage increases during transaction execution as the execution of a chaincode also takes place at the same time. The memory usage by the endorser is about $60\%$ during the idle phase and reaches to a maximum of $65\%$ during the chaincode execution.

%\begin{figure}[t]
%\centering
%\includegraphics[width=3in,height=2in]{img/HyperLedgerLAB.pdf}
%\captionsetup{justification=centering}
%\caption{Transaction latency with different endorsers (Lab)}
%\label{fig:comp_lab}
%\end{figure}

\subsection{Ethereum PoA Setup}
%\manos{TODO: Explain as part of the setup the one sealer with multiple copies and how this is similar to the HLF implementation}

In order to evaluate the Ethereum PoA platform, we construct a synthetic application as a cash (Ether) transfer application where Ether token is transferred from one account to another. We create two accounts i.e., source and the target account, and cash is transferred between accounts by calling \textit{sendTransaction}, an inbuilt function available in Ethereum implementation \textit{geth} to transfer funds (in Ether) between two accounts. For Ethereum, Proof of Authority consensus mechanism \textbf{Clique}, implemented in geth is used and experiments are performed with $1$, $2$ and $4$ sealers/validators. The results of each experiment are averaged over 5 independent runs. 

We deploy an Ethereum PoA etwork with a blocktime of $5$ seconds for our experiments as PoA consensus mechanism is more suitable to permissioned blockchain networks than the default PoW consensus mechanism. There are two kinds of nodes in a PoA network - \textit{Validators or Sealers}, who sign and create new blocks; - \textit{Non-Validators or Clients}, who do not have the authority to create new blocks and are mostly deployed in the network as interface for users to connect to blockchain network and submit transactions. We perform experiments in both lab and QMPSU for various configuration as listed below.

\begin{itemize}
    \item \textbf{Baseline-lab setup}: $2$ validator nodes co-located in the same host (Minix Device\footnote{http://minix.com.hk/}). Transactions are generated from within the host, and sent to one of the validators through Inter-process communication (IPC). Experiments are performed measuring transaction latency for $1$, $10$, $100$, $1000$ and $10000$ transactions fired in parallel.
    
    \item \textbf{Mesh-lab setup}: In order to evaluate the effect of the mesh, we perform the same experiments of Baseline-lab setup, but launching the transactions using WebSockets, from a powerful (Desktop machine) node that is located $2$ (short) mesh hops away from the Minix Device.
    
    \item \textbf{QMPSU setup}: We perform experiments in QMPSU in line with experiments performed for Hyperledger Fabric in QMPSU. To mimic single organisation scenario of Hyperledger Fabric, we authorise only one sealor/validator account and run multiple instances of the validator by varying number of validator instances from $1$ to $4$. We compare transaction latency for $100$ transactions fired in parallel when validator instances are placed randomly and when validator instances are placed in RPi3 nodes with the BASP heuristic. We also vary number of transactions from $1$ to $1000$ and record transaction latency. Apart from transaction latency, we measure CPU load and memory usage of validator and non-validator nodes when idle and busy.
\end{itemize}

\subsubsection{Ethereum PoA Experimental Results}

\begin{figure}[!h]
    \centering
    \begin{subfigure}{0.5\textwidth}
        \centering

        %old:EthereResultsBASE
        \includegraphics[width=3.1in,keepaspectratio]{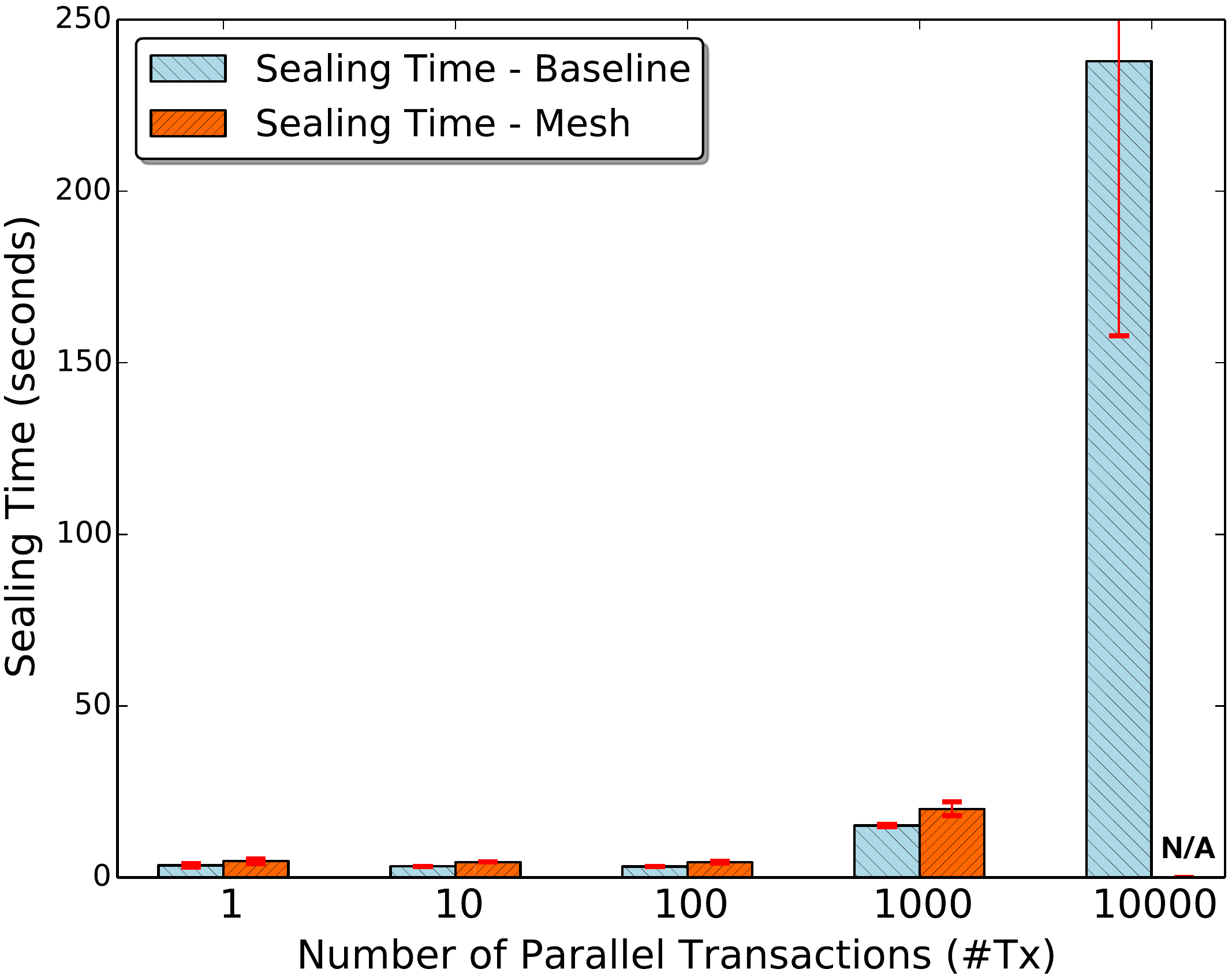}
        %\vspace{-.2cm}
        \caption{Baseline (client and miner on the same host)}
        \label{fig:labSealing}
    \end{subfigure}%
    \begin{subfigure}{0.52\textwidth}
        \centering
        %old: ethereumsresultMESH
        \includegraphics[width=3.1in,keepaspectratio]{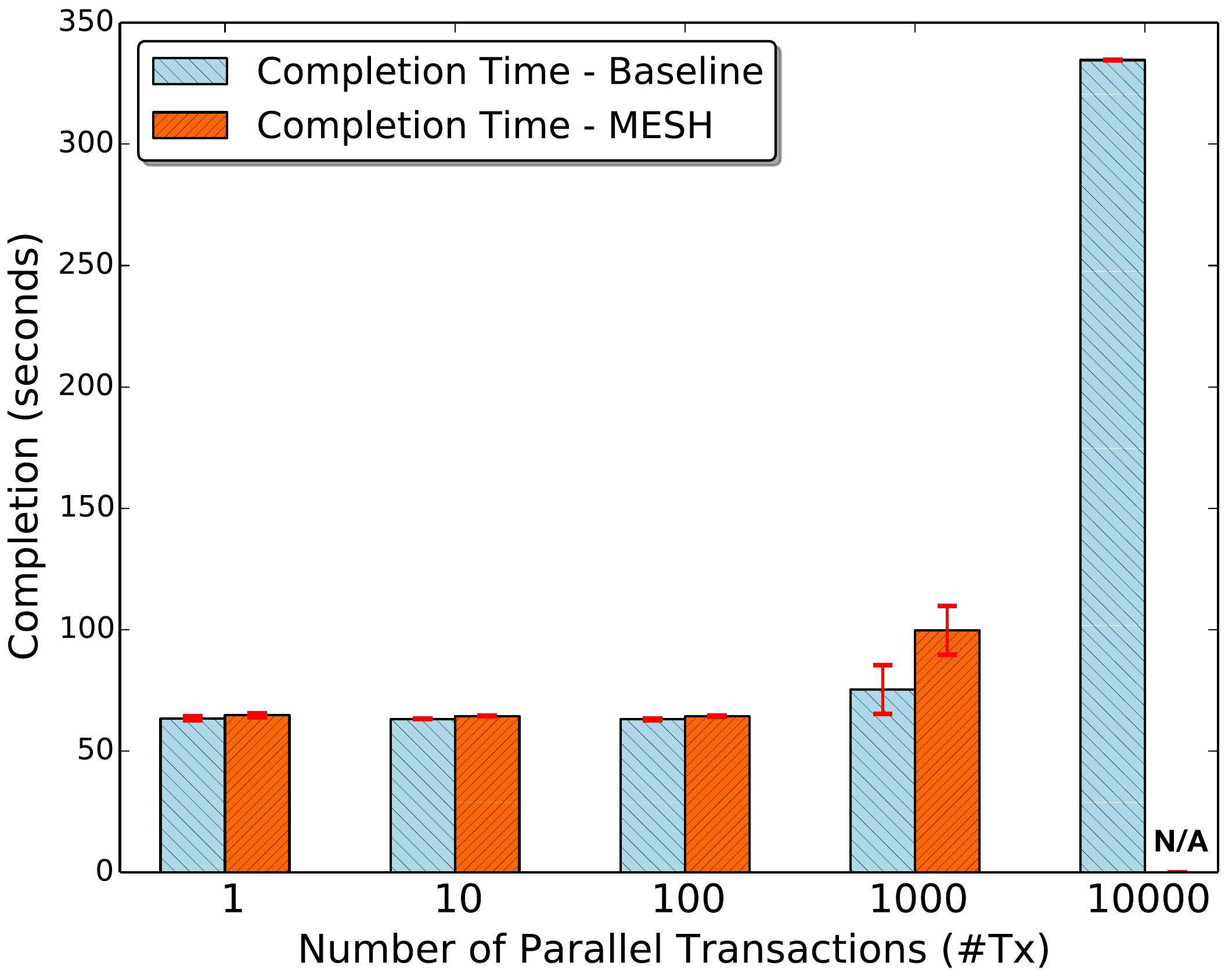}
        %\vspace{-.2cm}
        \caption{Mesh (miner and client 3-4 network hops away)}
        \label{fig:labCompletion}
    \end{subfigure}
    \caption{Sealing and Completion Time for Baseline/centralized and Mesh/distributed environment (Eth.PoA).}
    \label{fig:labmesh}
\end{figure}

\textbf{Transaction Latency:} In Ethereum, transaction latency is measured in multiple ways. Intuitively, transaction latency is the time between firing a transaction to the time it gets sealed. However, there is a significant probability that the mined block may not end up in the chain due to forking. Therefore, as mentioned earlier, it is a standard practice to consider confirmation of next $12$ blocks as finality for a transaction. In our experiments we measure both \textit{sealing time} and \textit{completion time}, see Figure \ref{fig:labmesh}, which we define as the time from firing of transaction upon receiving the $12$th confirmation, with the transaction under consideration being part of the chain. 

As an exploratory study, before jumping into QMPSU, we setup an Ethereum PoA private network following the Baseline-lab and Mesh-lab setups, measuring the sealing and completion time for $1$, $10$, $100$, $1000$ and $10000$ transactions. Figures \ref{fig:labSealing} and \ref{fig:labCompletion} compare the transaction latencies for the two different setups. 
%In baseline setup,, we have $2$ colocated validators, while in case of multinode setup in lab, the $2$ validators are located in two different physical devices. 
%
As expected, the Baseline setup shows lower latency than the Mesh one. Considering a blocktime of $5$ seconds, then in normal situations we expect a $sealingTime\in[0,blockTime]\equiv[0,5]$ and that $completionTime>=12*blockTime=60$. For both the sealing and completion time we observe that they show a normal behaviour for up to $100$ parallel transactions. 
%Moreover, we can observe that up to $1000$ transactions,  the baseline setup performs faster and better than the multinode setup. 
However, at $1000$ transactions, we already note increased delays in both the cases. It is interesting to point out here that an increase of $5$ seconds can be translated as a delay for the next block to be sealed. At $10000$ transactions, the multinode setup is completely saturated and does not respond. We observed that the majority of the transactions do not manage to get included in next $50$ blocks and are dropped, which in turn causes a timeout of the connector library, exiting with an exception. Even with the baseline setup, at $10000$ transactions, there is a large delay in getting the responses of block being sealed and confirmed. As a result, the rest of the experiments are performed with a maximum of $1000$ parallel transactions.

\begin{figure}[!h]
    \centering
    \begin{subfigure}{0.5\textwidth}
        \centering
        \includegraphics[width=3.1in,keepaspectratio]{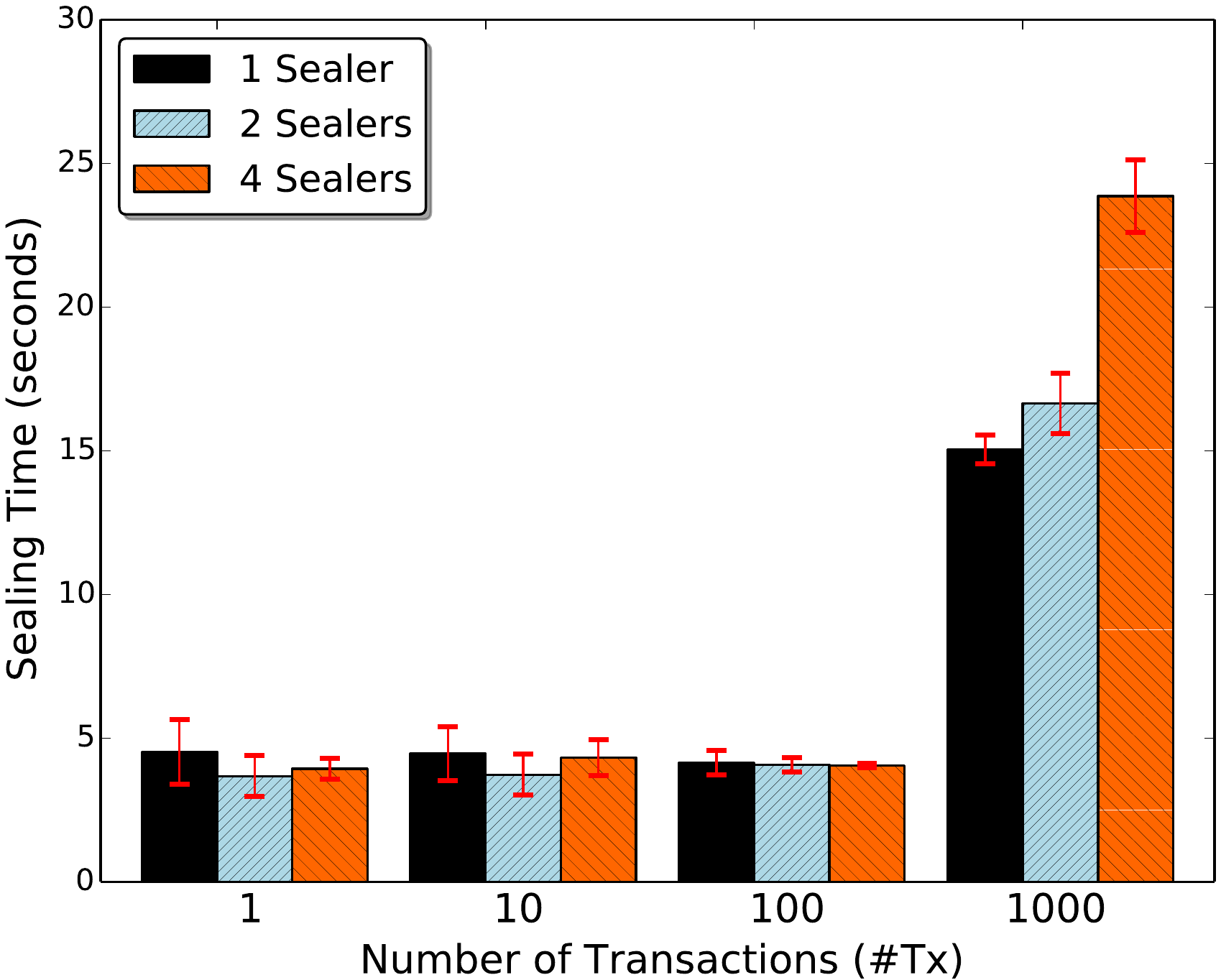}
        %\vspace{-.2cm}
        \caption{Completion time with 1, 2 and 4 Miners}
        \label{fig:multipleInstancesSeal}
    \end{subfigure}%
    \begin{subfigure}{0.52\textwidth}
        \centering
        \includegraphics[width=3.1in,keepaspectratio]{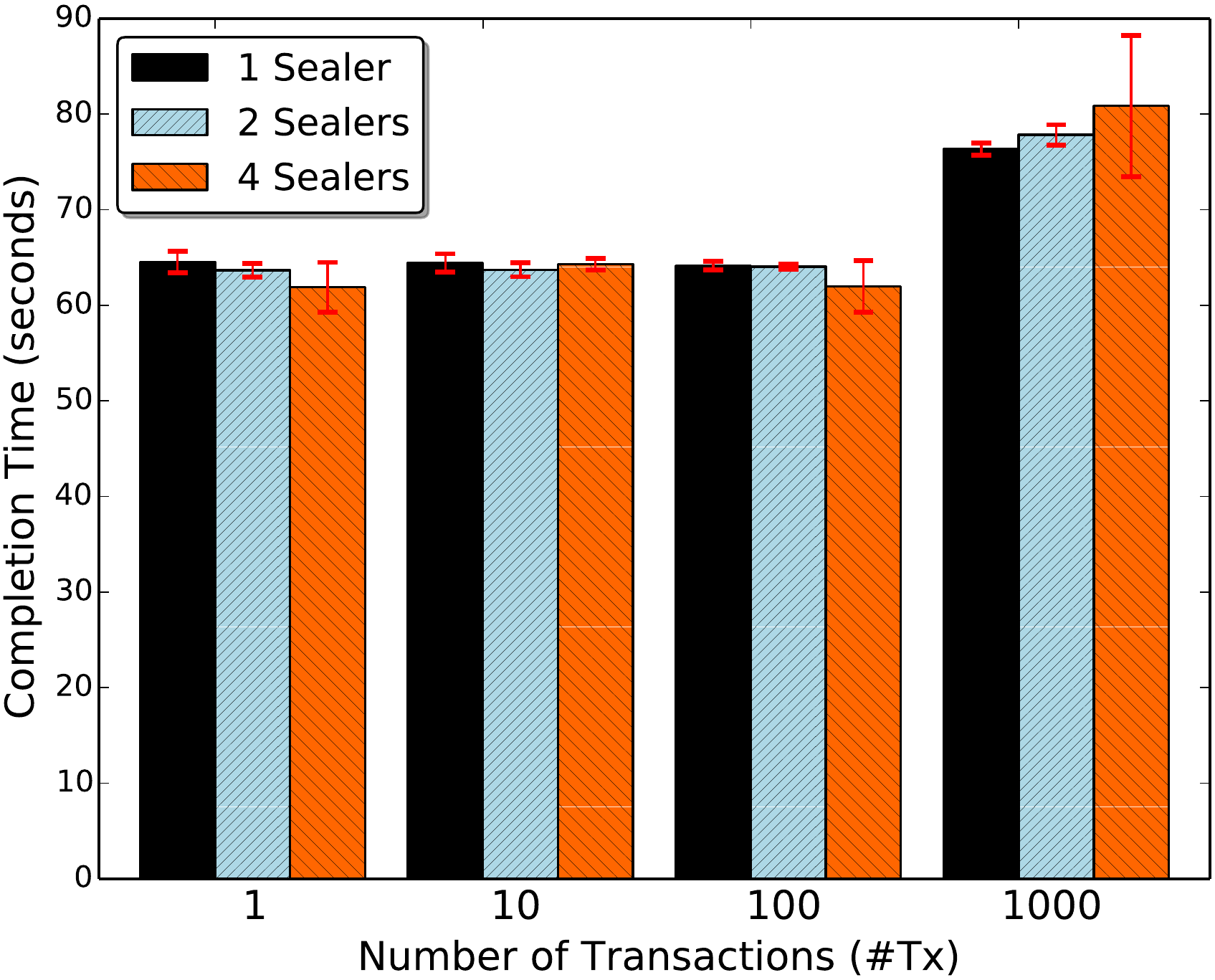}
        %\vspace{-.2cm}
        \caption{Mining time with 1, 2 and 4 Miners}
        \label{fig:multipleInstancesCompl}
    \end{subfigure}
    \caption{Sealing and Completion time for 1, 2 and 4 Sealers. 1, 10, 100 and 1000 transactions used (QMPSU, Eth.PoA).}
    \label{fig:bwtf}
\end{figure}

After the exploratory study, we performed experiments in QMPSU, obtaining the Figures  \ref{fig:multipleInstancesSeal} and \ref{fig:multipleInstancesCompl}, that show respectively the sealing and completion time for $1$, $10$, $100$ and $1000$ transactions fired in parallel with $1$, $2$ and $4$ validator instances. Considering a blocktime of $5$s and similarly to the exploratory experiments, Figure \ref{fig:multipleInstancesSeal} shows that up to $100$ transactions, all the transactions are verified and sealed in $1$ block. Beyond $100$ transactions, the number of blocks needed to accommodate all transactions fired increases beyond $3$ blocks. In the case of $1000$ transactions, that are accommodated in more than $3$ blocks, the total sealing time increases with more number of validator instances. The delay may be attributed to the latency generated by broadcasting the pending transactions to different validator nodes, once the current validator node reaches the block gas limit and cannot accommodate anymore transactions in the block. 
%Moreover, the latency in broadcast of sealed block from different validator nodes to the network may add up to this delay, considering very low blocktimes and big number of sealers. 
As far as the completion time is concerned, plotted in \ref{fig:multipleInstancesCompl}, we observe a similar behaviour to the sealing time. Between $1$ and $100$ parallel transactions fired, completion time is almost constant as empty blocks are sealed irrespective of number of transactions fired earlier, while shows an increased value for $1,000$ transactions. 

%2 Miners per node
%\begin{figure}[!h]
%    \centering
%    \begin{minipage}{0.5\textwidth}
%        \centering
%        \includegraphics[width=3.8in,keepaspectratio]{images/4Miners2perNode_CT.pdf}
%        \vspace{-.2cm}
%        \caption{Completion time with 1, 2 and 4 Miners (2 Miners)}
%        \label{fig:ether11}
%    \end{minipage}%
%    \begin{minipage}{0.52\textwidth}
%        \centering
%        \includegraphics[width=3.8in,keepaspectratio]{images/4Miners2PerNode_MT.pdf}
%        \vspace{-.2cm}
%        \caption{Mining time with 1, 2 and 4 Miners (2 Miners)}
%        \label{fig:tfd1}
%    \end{minipage}
%    %\caption{Bandwidth and traffic distribution in the qMp}
%    \label{fig:bwtf}
%\end{figure}

\begin{figure}[!h]
    \centering
    \begin{subfigure}{0.5\textwidth}
        \centering
        \includegraphics[width=3.1in,keepaspectratio]{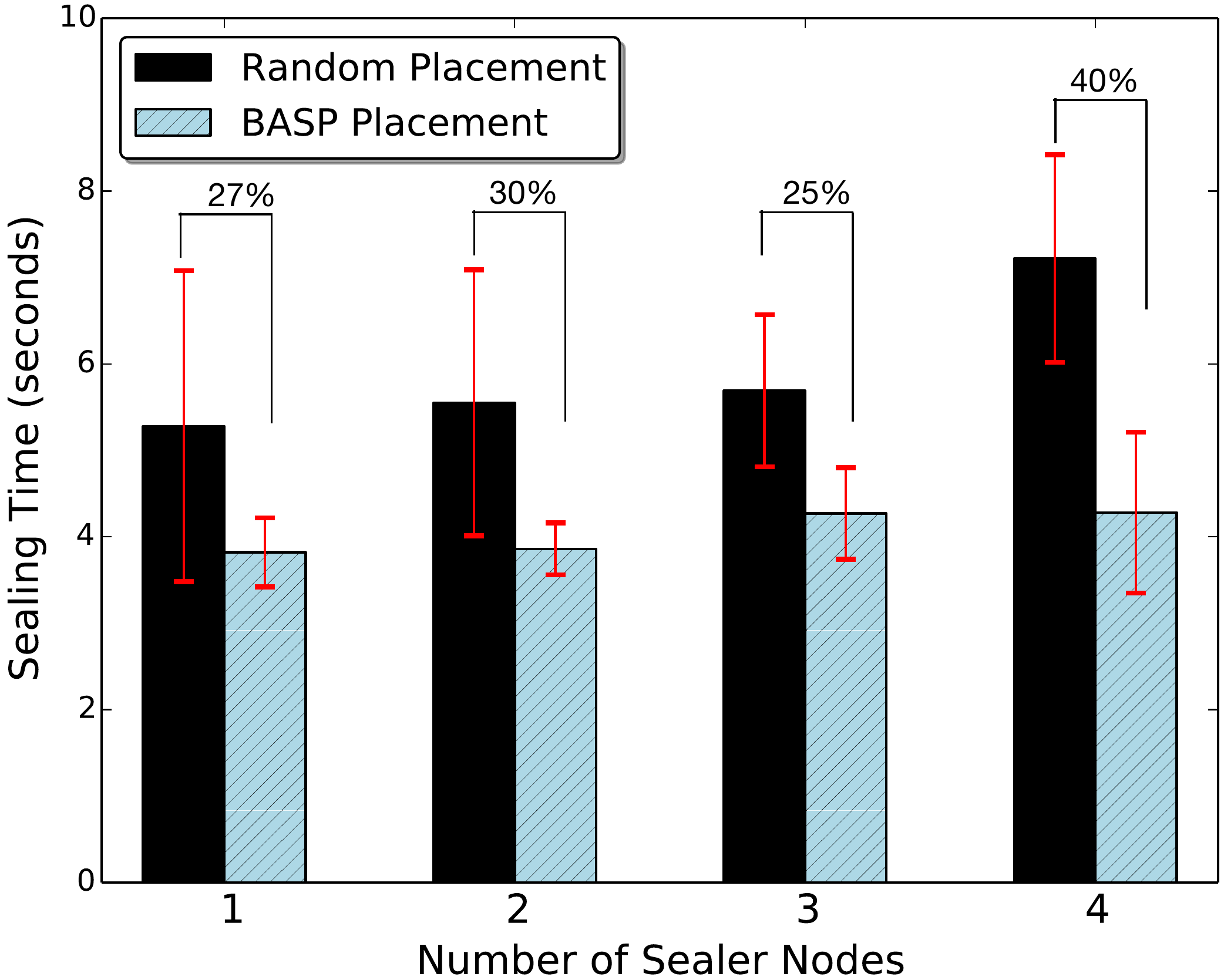}
        %old ethereummining time
        %\vspace{-.2cm}
        \caption{Sealing Time for 100 Tx (BEST vs Random)}
        \label{fig:baspST}
    \end{subfigure}%
    \begin{subfigure}{0.52\textwidth}
        \centering
        \includegraphics[width=3.1in,keepaspectratio]{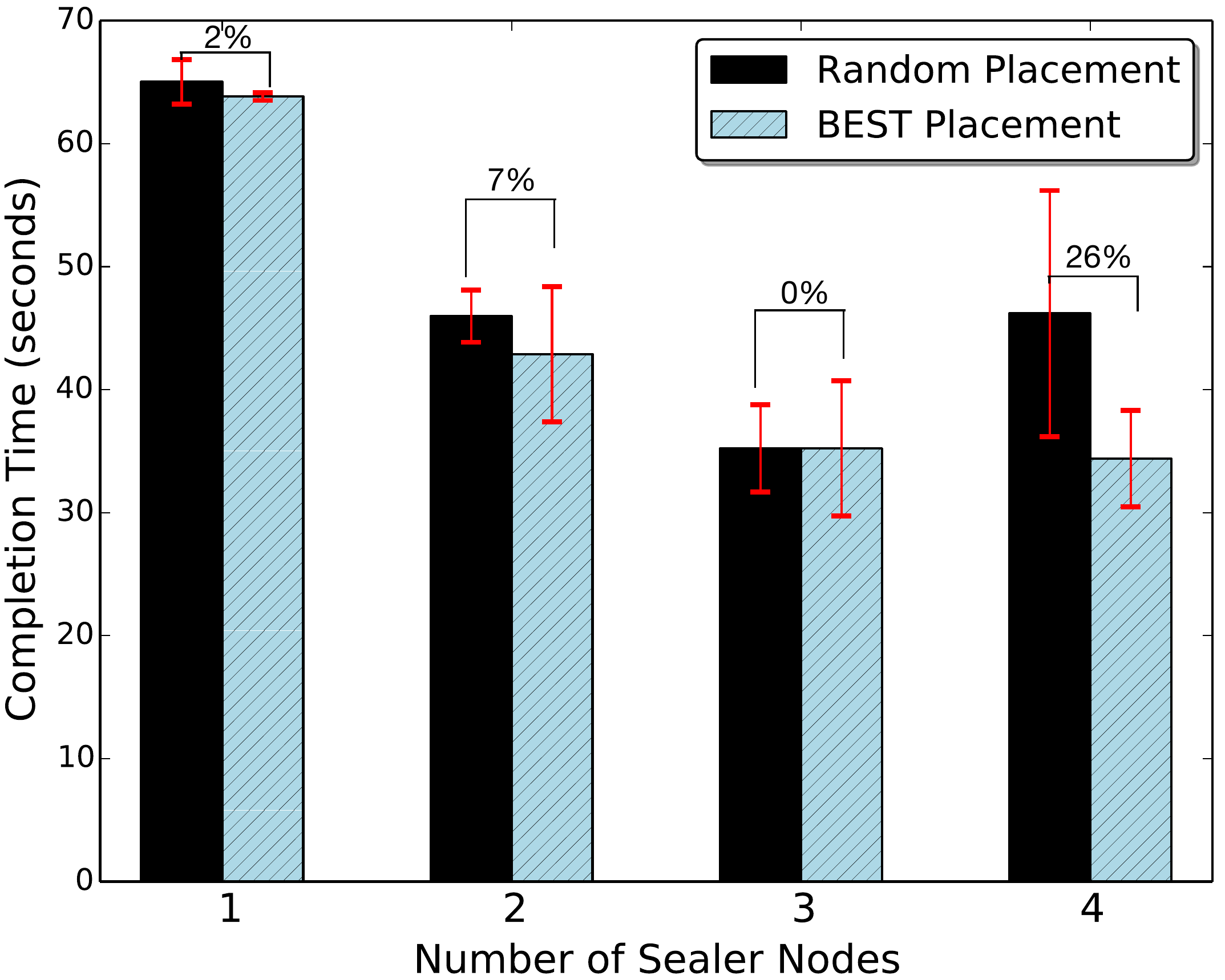}
        %\vspace{-.2cm}
        \caption{Completion time for 100Tx (BEST vs. Random)}
        \label{fig:baspCT}
    \end{subfigure}
    \caption{Sealing and Completion Time for 100 Transaction (BASP vs. Random) in (QMPSU, Eth.PoA)}
    \label{fig:baspR}
    %\caption{Mennansas as}
\end{figure}

\textbf{Placement:} Figure \ref{fig:baspST} and Figure \ref{fig:baspCT} depicts the sealing and completion time for different number of sealer nodes in the network, when nodes are placed randomly and with the BASP heuristic respectively. The figures reveal that BASP outperforms the Random placement when using up to $4$ sealer nodes. Moreover, as the number of sealer node increases, the gain tends to increase accordingly. For instance, when having up to $4$ sealer nodes, Figure \ref{fig:baspST} shows that the gain brought by BASP over random is $3$ seconds which is $40$\% improvement. The same thing happens with completion time in Figure \ref{fig:baspCT}, where the gain brought by BASP over random is $12$ seconds which is $26$\% of improvement. 

These figures demonstrate the importance of the sealer node location in the network. In a challenging environment such as wireless mesh network, the placement heuristics that are agnostic to the state of the underlying network may lead to important inefficiencies. Our result demonstrates that placement of sealer nodes can become even more crucial when number of transaction is higher (e.g., 1,000, 10,000 etc).

\begin{figure}[t]
\centering
\includegraphics[width=5in,keepaspectratio]{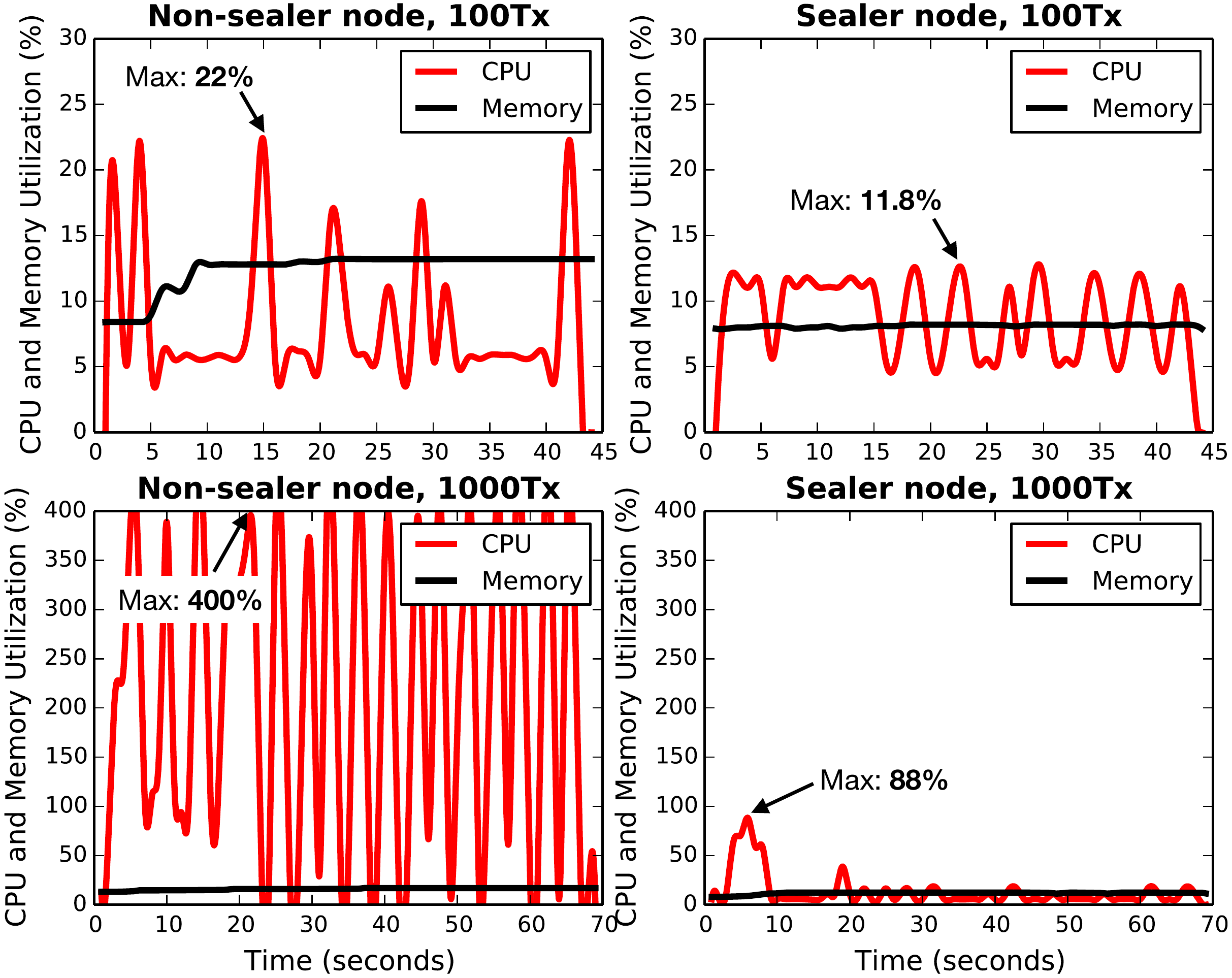}
\captionsetup{justification=centering}
\caption{CPU and Memory utilization with 100 and 1000 transactions (Eth.PoA)}
\label{fig:cpumemory}
\end{figure}

\textbf{Resource Consumption:} In order to understand the resource usage of the participating nodes, we measured CPU Load and memory usage in sealer and non-sealer nodes when $100$ and $1000$ transactions are fired, as plotted in Figure \ref{fig:cpumemory}. We record the measurement from the time we fire transactions to the time the transactions are considered completed (i.e, $12$ confirmations in Ethereum). The procedure we follow here is that, we fire transactions to  non-sealer node; non-sealer node broadcasts the transactions to sealer node where it is sealed. We follow this strategy to get an idea of load generated by independent processes like accepting transactions and sealing. We observe that when $1000$ transactions are fired in parallel, the non-sealer node is saturated heavily in all the $4$ cores of the RPi3 board. Even with $100$ transactions, the CPU load on non-sealer node is pretty high. While the validation process loads the CPU only moderately for both $100$ and $1000$ transactions. It is also demonstrated that the non-sealer has higher memory consumption compared to the sealer node in both cases. However, the maximum memory usage is still below $20$\% and is not a bottleneck in the blockchain network. As expected, the sealing process, following a PoA scheme, is not demanding in terms of resources. On the other hand, accepting transactions and forwarding them to the sealer nodes, as the non-sealer node of our experiment does, seems to be very resourceful, and can saturate the node. 

%\begin{figure}[t]
%\centering
%5\includegraphics[width=5in,keepaspectratio]{images/distributionETHER.png}
%\captionsetup{justification=centering}
%\caption{ECDF of completion time}
%\label{fig:cpu_memory}
%\end{figure}

\subsection{Discussion}

%\begin{description}
\textbf{Hyperledger Fabric}: As we observed in our experiments, in terms of resource consumption, the endorser nodes can prove to be a bottleneck. We believe that this bottleneck is because of the execution of an additional chaincode container at each endorsing node. In our current study we only considered one endorser node to study the resource utilization with a simple endorsement policy encoded in the corresponding chaincode. It might get more complicated when we consider more than one endorser, and more sophisticated endorsement policies. However, as discussed in the paper, if done right it can actually improve performance. In addition to this, the actual distribution of endorsing peers in a production network, such as QMPSU, might also affect the network performance (both in terms of CPU utilization and transaction latency). Therefore we advise caution when in designing an endorsing policy that is also cognizant of the underlying network infrastructure (i.e, topology, capacity, performance, etc), especially in the resource constrained nature of CMNs. A deployment strategy and an apt endorsement policy balancing the load on various endorsers in the network can improve the performance of the blockchain network and allow scaling of the blockchain network without forming a bottleneck.  As far as the orderers are concerned, horizontal scaling by adding more nodes is possible, nevertheless, this would need some sort of mechanism for syncing between instances. For instance, it is possible to have multiple instances of ordering service nodes all connected to a single fault tolerant service (Kafka) that would do the ordering (crash fault tolerant).

%\manos{Write about horizontal scaling by increasing number of orderers? tradeoff with the consensus communication between the orderers}
%\manos{Write that this results apply for value transfer. Execution of chaincode is different, since a new docker is spawned for each chaincode.}

\textbf {Ethereum PoA}:The results presented earlier concerning Ethereum show that it can be used successfully as private permissioned blockchain in a mesh environment, using PoA consensus. Nonetheless, there are various parameters to be adjusted and bottlenecks that need to be discussed. Unlike HLF, in Ethereum PoA there is no clear horizontal scaling pattern. While having a lot of sealers could balance the incoming transactions, the transaction throughput is largely affected by the hardware resources like CPU and memory of the nodes who accept the transactions and less affected by the number of nodes . This, depending on the frequency of transactions generated, can be a significant issue for mesh like environments, since the hardware used is usually low-power/low-cost devices. Moreover, the broadcasting of the pending transactions between the sealers can become problematic over non-stable mesh connections, especially between remote nodes, or nodes connected with lossy links. This situation could also deteriorate by an increased number of nodes and small blocktimes, leading to higher frequency and higher number of message exchanges between the sealers. On the other hand, these effects could be moderated by utilising smart placement algorithms like BASP, which would play a significant role in avoiding network saturation, by placing the sealers in locations that would minimise the overhead of the blockchain. Finally, while we deploy multiple clones of one sealer, other approaches are possible, like having multiple sealer accounts, considering that a minimum of $N/2+1$ instances of them are always available~\citep{EIP225}.\\

Despite some resource inefficiencies in the execution of chaincodes and smart contracts, we anticipate a gradual improvement in future iterations of both software platforms in this regard. We, however, consider the observed performance and latency (in the range of 1-2 minutes) acceptable  for the problem at hand, which is of balancing and transferring payments from consumers to providers that in practice can take from several minutes, hours or even days for every connectivity consumer and provider. Further, we believe that in the case of community mesh networks a decentralized and trustless environment in conjunction with tamper-evident record keeping in the form of a distributed ledger takes preference over performance. This is so that in order for these networks to function seamlessly and in a mostly peer-to-peer but conflict-free scenario. Contrary to public and open blockchains, private and permissioned distributed ledgers for this purpose have completely different scalability ranges. As described in Section \ref{sec:qmp} and \citep{MACCARI2015175}, typical mesh access networks cover a city or neighbourhood, share one or two Internet gateways, and have in the range of or below two hundred routers, with a relatively small number of servers where blockchain processes can execute, in several servers provided by different participants to ensure trust and reliability. %Combined with economic compensations done in the range of hours to days
It becomes clear that automation and irreversibility guarantees provided by blockchain brings down the latency of economic settlements from days and hours to minuetes and seconds.  Therefore we are of the view that economic compensation in the particular case of community networks takes preference over scalability which is not the immediate pressing issue.

\section{Related Work}
 \label{sec:relwork}

There are works that explore aspects of how mesh networks can be combined with blockchain technologies to provide connectivity under an economic model in a decentralised manner involving the independent providers and consumers of a crowdsourced ISP.  In contrast to most of the works mentioned in this section, we specifically consider the implications of deploying the blockchain paradigm to a, still in use, production environment such as that of CMNs.

\textbf{Comparison of Public and Private Blockchain Platforms:}
The work of Suporn et al. \citep{PerformanceBlock} presents performance analysis of Hyperledger Fabric and Ethereum as private blockchain platforms with varying number of transactions. They conduct their experiments in Amazon AWS EC2 instances. Their assessment shows that Hyperledger Fabric consistently outperforms Ethereum across all evaluation metrics such as execution time, latency and throughput. Further, they claim that both platforms are still not competitive with current database systems in terms of performance when using high workloads. The work in \citep{BlockSurvey} discusses various consensus protocols used in blockchain and comparative analysis of Hyperledger Fabric and Ethereum. The study \citep{SchererThesis} compares the public blockchain with permissioned blockchain and discusses the trade-offs among decentralization, scalability and security in the two approaches. Sousa et al. \citep{Bessani2017} present the design, implementation and evaluation of a BFT ordering service for Hyperledger Fabric based on the the BFT-SMART state machine replication/consensus library. Their results show that Hyperledger Fabric with their ordering service can achieve up to ten thousand transactions per second and write a transaction irrevocably in the blockchain in half a second, even with peers distributed over different continents. The Blockbench \citep{Blockbench} is a framework for analyzing private blockchains. It serves as a fair means of comparison for different platforms and enables deeper understanding of different system design choices. They use Blockbench to conduct comprehensive evaluation of three major private blockchains: Ethereum, Parity\footnote{\url{https://www.parity.io/}} and Hyperledger Fabric. Their results demonstrate that these systems are still far from replacing the current database systems in traditional data processing workloads. In contrast to most of the works mentioned in this section, we specifically consider the implications of deploying the blockchain paradigm to a, still in use, production environment such as that of CMNs.

\textbf{Payments in Mesh Networks using Blockchain:}
There are several projects in development that combine the payment in mesh networks with blockchain. Althea Mesh \cite{Althea} provides last-mile connectivity for the Internet access. Althea allows routers to pay each other for bandwidth using cryptocurrency payment channels. Nodes only pay neighbours for forwarding packets. RightMesh \cite{RightMesh} is a software-based, ad hoc mobile mesh networking platform and protocol using blockchain technology and RMESH tokens. RightMesh integrates into the Ethereum blockchain to provide unique identities for each node in the mesh. AmmbrTech \cite{AmmbrTech} develops solutions that combine networking devices and software tools around a combination of digital identity, 
local and global blockchain and distributed ledgers, wireless mesh networks, and artificial intelligence to self-adapt the system.

\textbf{Off-Chain Payment Networks:} Payment channels allow to establish a direct peer-to-peer payment between two parties such that their individual transactions are not required to be written to the blockchain. The Lightning Network (LN) \cite{LightningNetwork} is a layer-2 protocol built on top of the current Bitcoin protocol. The idea of such a protocol is to deploy an overlay network (using payment channels) where off-chain payments can be made from node to node on a path of such an overlay without trusting any of the nodes in the path. The work in \cite{LN2} proposes a more general technique for nodes to apply fees for forwarding payments in LNs. Further, they propose a multipath routing payment scheme which is able to significantly reduce the fees paid by users. Revive \cite{Revive} is a rebalancing scheme for payment channels that allows a user to utilize any other of his channels for rebalancing a particular channel. The Raiden Network \cite{Raiden} is an off-chain scaling solution, enabling near-instant, low-fee and scalable payments in Ethereum. Sprites \cite{sprites} inspired by Raiden and Lightning Network, aims to minimize the worst-case collateral costs of indirect off-chain payments. Bolt \cite{bolt} is a protocol that constructs anonymous payment channels between two mutually distrustful parties. Their approach allow for secure, instantaneous and private payments that substantially reduce the storage burden on the payment network.

\textbf{Resource trading:} Several studies provide economic analysis and designs for resource trading. Route Bazaar \cite{IgnacioHotOS} is a backward-compatible system for flexible Internet connectivity. Inspired by the decentralised construction of trust in cryptocurrencies, Route Bazaar uses a decentralised public ledger and cryptography to provide Autonomous Systems (ASes) with automatic means to form, establish, and verify end-to-end connectivity agreements. Tycoon~\cite{Lai:2005} is a market based distributed resource allocation system based on proportional share that uses auctions for resources, such as computing, storage or network traffic, that uses a centralised bank component based on digitally signed receipts to attest payments, that can be used to claim access and usage of resources later on. Request Network \cite{RequestNetwork} is a decentralised network that allows anyone to request a payment (a Request Invoice) for which the recipient can pay in a secure way. All of the information is stored in a decentralised authentic ledger. Request can be seen as a layer on top of Ethereum which allows requests for payments that satisfy a legal framework.

%Most of the above mentioned works are not done in CMNs context and are not applicable to our scenario.
\section{Conclusion}
\label{sec:conclusion}

The missing ingredient for widespread adoption of decentralized access networks (such as community mesh access networks) has always been the issue of economic sustainability.  In this paper, we take on the issue of addressing trustworthy economic sustainability by proposing the need for an economic substrate built using blockchain that can keep a record of the transactions that keeps track of the contributions (of nodes, links, Internet gateways, maintenance), consumption of communication network's resources as its economic compensation in a transparent, decentralized and trusted manner.
The evaluation of the Hyperledger Fabric and Ethereum blockchain deployment in a real test network in a laboratory setup and a real production mesh network gives us an understanding of the performance, overhead, influence of the underlying network, and limitations of the two platforms. The results show critical aspects that can be optimized in a Hyperledger Fabric and an Ethereum PoA local blockchain deployment, in the perspective of decentralized networks, where several components can prove to be bottlenecks and therefore put a limiting effect on the rate of economic transactions in a mesh network.
Future work will expand the evaluation to a wider range of hardware and network configurations considering real and synthetic transaction traces. We will also consider the influence of the execution of non-trivial smart contracts, with a more realistic design of an endorsement policy (in the form of chaincode(s)).

\section*{Acknowledgments}
This paper has been supported by the AmmbrTech Group, the Spanish government TIN2016-77836-C2-2-R and the European Community H2020 Programme netCommons (H2020-688768). The authors would like to thank the people from the Guifi.net (Guifi-Sants) community network for hosting the servers and supporting the experiments.

\bibliographystyle{WileyNJD-AMA}
\bibliography{bibliography/references.bib}

\begin{thebibliography}{10}
\providecommand \doibase [0]{http://dx.doi.org/}%

\bibitem{akyildiz2005wireless}
Akyildiz IF, Wang X, Wang W. Wireless mesh networks: a survey. {\it Computer
  networks} 2005\string; 47(4)\string: 445--487.

\bibitem{baig2016making}
Baig R, Dalmau L, Roca R, Navarro L, Freitag F, Sathiaseelan A. Making
  community networks economically sustainable, the guifi. net experience. In:
  ACM. ; 2016\string: 31--36.

\bibitem{ITU}
ITU . {International Telecommunications Union, ICT Facts and Figures 2016}.
  http://www.itu.int/en/ITU-D/Statistics/Documents/facts/ICTFactsFigures2016.pdf;
  2016.

\bibitem{nakamoto2008bitcoin}
Nakamoto S. Bitcoin: A peer-to-peer electronic cash system. 2008.

\bibitem{Neumann:2016}
Neumann A, L{\'o}pez E, Cerd{\`a}-Alabern L, Navarro L. Securely-entrusted
  multi-topology routing for community networks. In: IEEE. IEEE; 2016.

\bibitem{Selimi2018CryBlock}
Selimi M, Kabbinale AR, Ali A, Navarro L, Sathiaseelan A. Towards
  Blockchain-enabled Wireless Mesh Networks. In: CryBlock'18. ACM; 2018; New
  York, NY, USA\string: 13--18

\bibitem{wood2014ethereum}
Wood G. Ethereum: A secure decentralised generalised transaction ledger. {\it
  Ethereum Project Yellow Paper} 2014\string; 151\string: 1--32.

\bibitem{Hyper2018}
{Androulaki} E, others . {Hyperledger Fabric: A Distributed Operating System
  for Permissioned Blockchains}. {\it ArXiv e-prints} 2018.

\bibitem{Bessani2017}
{Sousa} J, {Bessani} A, {Vukoli{\'c}} M. {A Byzantine Fault-Tolerant Ordering
  Service for the Hyperledger Fabric Blockchain Platform}. {\it ArXiv e-prints}
  2017.

\bibitem{LlorencMSWiM}
Cerd\`{a}-Alabern L, Neumann A, Escrich P. Experimental Evaluation of a
  Wireless Community Mesh Network. In: MSWiM '13. ACM; 2013; New York, NY,
  USA\string: 23--30

\bibitem{Baig:2018}
Baig R, Freitag F, Navarro L. Cloudy in guifi.net: Establishing and sustaining
  a community cloud as open commons. {\it Future Generation Computer Systems}
  2018.
\newblock \href {\doibase https://doi.org/10.1016/j.future.2017.12.017} {doi:
  https://doi.org/10.1016/j.future.2017.12.017}

\bibitem{Selimi2018}
Selimi M, Cerd{\`a}-Alabern L, Freitag F, Veiga L, Sathiaseelan A, Crowcroft J.
  A Lightweight Service Placement Approach for Community Network Micro-Clouds.
  {\it Journal of Grid Computing} 2018.
\newblock \href {\doibase 10.1007/s10723-018-9437-3} {doi:
  10.1007/s10723-018-9437-3}

\bibitem{Gelly2018}
Coimbra ME, Selimi M, Francisco AP, Freitag F, Veiga L. Gelly-Scheduling:
  Distributed Graph Processing for Service Placement in Community Networks. In:
  ACM; 2018.

\bibitem{SelimiCCGrid}
Selimi M, Cerda-Alabern L, Sanchez-Artigas M, Freitag F, Veiga L. Practical
  Service Placement Approach for Microservices Architecture. In: ; 2017\string:
  401-410

\bibitem{weber2017availability}
Weber I, Gramoli V, Ponomarev A, et al. On availability for blockchain-based
  systems. In: IEEE. ; 2017\string: 64--73.

\bibitem{EIP225}
Developers E. Clique PoA protocol \& Rinkeby PoA testnet.
  \url{https://github.com/ethereum/EIPs/issues/225};  2018.

\bibitem{MACCARI2015175}
Maccari L, Cigno RL. A week in the life of three large Wireless Community
  Networks. {\it Ad Hoc Networks} 2015\string; 24\string: 175 - 190.
\newblock Modeling and Performance Evaluation of Wireless Ad-Hoc Networks\href
  {\doibase https://doi.org/10.1016/j.adhoc.2014.07.016} {doi:
  https://doi.org/10.1016/j.adhoc.2014.07.016}

\bibitem{PerformanceBlock}
Pongnumkul S, Siripanpornchana C, Thajchayapong S. Performance Analysis of
  Private Blockchain Platforms in Varying Workloads. In: ; 2017\string: 1-6

\bibitem{BlockSurvey}
P. S, M. S, Sethumadhavan M. On Blockchain Applications: Hyperledger Fabric And
  Ethereum. {\it International Journal of Pure and Applied Mathematics} 2018.

\bibitem{SchererThesis}
Scherer M. Performance and Scalability of Blockchain Networks and Smart
  Contracts. Master's thesis. Umea University, Department of Computing Science.
   2017.

\bibitem{Blockbench}
Dinh TTA, others . BLOCKBENCH: A Framework for Analyzing Private Blockchains.
  In: SIGMOD '17. ACM; 2017; New York, NY, USA\string: 1085--1100

\bibitem{Althea}
{Althea: An incentivized mesh network protocol}.
  https://altheamesh.com/documents/whitepaper.pdf; .

\bibitem{RightMesh}
{RightMesh}. https://www.rightmesh.io/docs/RightMesh\_TWP5.pdf; .

\bibitem{AmmbrTech}
{Ammbr: Blockchain-enbaled Wireless Mesh Networks}. https://ammbrtech.com/; .

\bibitem{LightningNetwork}
Poon J, Dryja T. {The Bitcoin Lightning Network: Scalable Off-Chain Instant
  Payments}. https://lightning.network/lightning-network-paper.pdf; .

\bibitem{LN2}
Stasi GD, Avallone S, Canonico R, Ventre G. Routing payments on the Lightning
  Network. In: IEEE; 2018.

\bibitem{Revive}
Khalil R, Gervais A. Revive: Rebalancing Off-Blockchain Payment Networks. In:
  CCS '17. ACM; 2017; New York, NY, USA\string: 439--453

\bibitem{Raiden}
{The Raiden Network}. https://raiden.network/; .

\bibitem{sprites}
Miller A, Bentov I, Kumaresan R, McCorry P. Sprites: Payment Channels that Go
  Faster than Lightning. {\it CoRR} 2017\string; abs/1702.05812.

\bibitem{bolt}
Green M, Miers I. Bolt: Anonymous Payment Channels for Decentralized
  Currencies. In: CCS '17. ACM; 2017; New York, NY, USA\string: 473--489

\bibitem{IgnacioHotOS}
Castro I, Panda A, Raghavan B, Shenker S, Gorinsky S. Route Bazaar: Automatic
  Interdomain Contract Negotiation. In: {USENIX} Association; 2015; Kartause
  Ittingen, Switzerland.

\bibitem{Lai:2005}
Lai K, Rasmusson L, Adar E, Zhang L, Huberman BA. Tycoon: An Implementation of
  a Distributed, Market-based Resource Allocation System. {\it Multiagent Grid
  Syst.} 2005\string; 1(3)\string: 169--182.

\bibitem{RequestNetwork}
{Request Network: A decentralized network for payment requests}.
  https://request.network/assets/pdf/request\_whitepaper.pdf; .

\end{thebibliography}

%\input{bib.tex} %%% Bibliography from bbl file
%\nocite{*}% Show all bib entries - both cited and uncited; comment this line to view only cited bib entries;
%\bibliography{bibliography/references}%
\end{document}